\documentclass[a4paper,11pt]{article}

\usepackage{jheppub}
\usepackage[T1]{fontenc}
\usepackage{graphicx}% Include figure files
\usepackage{hyperref}
\usepackage{amsmath}
\usepackage{epsfig}
\usepackage{subfig}
\usepackage{xcolor}
\usepackage{natbib}
\usepackage{bm}% bold math

\def\be{\begin{equation}}
\def\ee{\end{equation}}
\def\bea{\begin{eqnarray}}
\def\eea{\end{eqnarray}}
\def\NO{\nonumber}
\def\gev{\mathrm{~GeV}}

\title{Kinematic distributions of the $\eta_c$ leptoproduction in association with light hadrons}% Force line breaks with \\

\author[a]{Hong-Fei Zhang,}
\author[a]{Xue-Mei Mo}
\affiliation[a]{Guizhou University of Finance and Economics, Guiyang, 550025, China}
\emailAdd{shckm2686@163.com}
\emailAdd{dmtmmu@163.com}

\abstract{
The $\eta_c$ meson leptoproduction is calculated within the nonrelativistic QCD framework for the first time.
It is found that the colour-singlet channel, although suppressed by a factor of $\alpha_s$ relative to the colour-octet ones,
provides the important contribution for almost all the experimental conditions,
which disagrees with some of the expectations before computation.
We present the differential cross sections with respect to $p_t^2$, $p_t^{\star2}$, $Q^2$, $W$, and $z$,
for both HERA and EIC experimental conditions as a reference for future studies.
The scale dependence and long-distance-matrix-element dependence are also investigated in this paper.
}
\keywords{Deep inelastic scattering, Nonrelativistic QCD}
\begin{document}

\maketitle
\bibliographystyle{JHEP}

\section{Introduction\label{sec:introduction}}

The quarkonium production mechanism is a puzzle hanging for decades.
In 1994, the colour-octet mechanism~\cite{Bodwin:1994jh} was proposed to explain the significant discrepancy
between theory and the measurement of the $J/\psi$ hadroproduction carried out at the Tevatron~\cite{CDF:1992cmg}.
Although this discrepancy was mended by an extremely large short-distance coefficient of the $c\bar{c}(^3S_1^{[8]})$ hadroproduction~\cite{Braaten:1994vv},
some new troubles emerge and further efforts were devoted to both theoretical computations and experimental measurements.
Among the difficulties, a well-known one is the $J/\psi$ polarization puzzle,
i.e. the Tevatron measurement~\cite{CDF:2007msx} of the $J/\psi$ polarization contradicts the $^3S_1^{[8]}$-dominance picture,
in which most of the $J/\psi$'s are produced through one-gluon fragmentation and thus the longitudinal polarization is greatly suppressed.
It was not until the next-to-leading order calculations was achieved did physicists find possibilities to solve this puzzle.
The behavior of the $^3P_J^{[8]}$ short-distance coefficients are completely changed at QCD next-to-leading order~\cite{Ma:2010yw, Butenschoen:2010rq, Ma:2010jj, Butenschoen:2011yh},
which leads to new values of the long-distance matrix elements that disfavor the $^3S_1^{[8]}$ dominance picture.
Three groups calculated the QCD corrections to the $J/\psi$ hadroproduction and studied its polarization~\cite{Butenschoen:2012px, Chao:2012iv, Gong:2012ug},
however, with different fitting strategies, they got completely different values of the long-distance matrix elements,
and their perspectives on the $J/\psi$ polarization puzzle were also different.
As pointed out in Reference~\cite{Sun:2015pia}, the $J/\psi$ production data alone can not fix all the three long-distance matrix elements.
A milestone was the measurement of the $\eta_c$ hadroproduction~\cite{Barsuk:2012ic, Aaij:2014bga},
which, provides unique information for the charmonia production, and, by virtue of heavy quark spin symmetry,
can help to eliminate the degrees of freedom in the fit of the long-distance matrix elements for the $J/\psi$ production~\cite{Han:2014jya, Zhang:2014ybe}.
It is demonstrated in Reference~\cite{Sun:2015pia} that the $J/\psi$ and $\eta_c$ hadroproduction data along with the $J/\psi$ polarization data at hadron colliders
can be well described with one set of long-distance matrix elements.
Unfortunately, the theoretical results using this set of parameters disagree with the Belle data~\cite{Belle:2009bxr, Li:2014fya};
there does not exist any set of long-distance matrix elements that can describe all the existing data.

Inspired by the fact that the QCD corrections to quarkonium production processes are usually remarkable,
some people expect that the next-to-next-to-leading-order correction to the colour-singlet channel might also be important~\cite{Haberzettl:2007kj},
and thus the colour-octet long-distance matrix elements might have been overestimated.
They sought the possibility to interpret the existing data in the absence of the colour-octet contributions
and found that the experimental data of some processes,
e.g. the $J/\psi$ production at B-factories~\cite{Ma:2008gq, Gong:2009kp, Gong:2009ng},
is saturated by the colour-singlet channel.
In order to clarify this puzzling problem,
it is important to find out more processes of which the colour-singlet contributions are negligible compared to the colour-octet ones.
In Reference~\cite{Zhang:2019wxo}, the authors studied the $\eta_c$ photoproduction in electron-proton collisions and found that,
in the direct photon processes, the colour-singlet channel is 1-2 orders of magnitude smaller than the colour-octet ones.
However, this feature is ruined by the resolved-photon contributions which enhance the colour-singlet part by two orders of magnitude.

In this paper, we study the $\eta_c$ production in deep inelastic electron-proton ($ep$) scattering.
In this process, the resolved-photon contributions are negligible, at the same time, the good feature,
namely the colour-singlet $^1S_0$ $c\bar{c}$ is produced with at least two gluons emitted, is retained.
Although the short-distance coefficient for the colour-singlet process is suppressed by a factor of $\alpha_s$ comparing with the colour-octet ones,
the colour-singlet long-distance matrix element is enhanced by some powers of $1/v$.
Since for the charmonia, $v^4$ is approximated to be 0.1,
one cannot naively conclude a colour-octet dominance picture in the $\eta_c$ photoproduction just from the analysis of the scaling.
This process was first investigated in Reference~\cite{Hao:2000ci}.
However, as pointed out in References~\cite{Zhang:2017dia, Sun:2017nly}, there might be some mistakes in the calculations of the quarkonium leptoproduction at that time,
and in any case, the results for the colour-singlet channel was not given in Reference~\cite{Hao:2000ci}.
As a matter of fact, this process is quite complicated due to the nonzero invariant mass of the initial photon,
which can be seen from the fact that the first correct complete nonrelativistic-QCD result for the $J/\psi$ leptoproduction
at QCD leading order was obtained in Reference~\cite{Sun:2017nly} in as late as 2017;
the first QCD next-to-leading-order result of the colour-singlet $J/\psi$ production was completed in 2017~\cite{Sun:2017wxk},
while the paralelle results for the $J/\psi$ photoproduction was published more than ten years ago~\cite{Artoisenet:2009xh};
and its colour-octet counterparts are still lacking.
This paper will, for the first time, present the complete results of the $\eta_c$ leptoproduction within the nonrelativistic QCD framework,
and discuss the possibility of distinguishing the colour-singlet and colour-octet mechanisms using this process.
The experimental data can be obtained by revisiting of the HERA data.
Further, we suggest it be measured in the future Electron-Ion Collider (EIC).

In Section~\ref{sec:framework} we outline the calculation framework.
Having the essential analytical formulas, we present the numerical results in Section~\ref{sec:results}.
A concluding remark is presented in the last section.

\section{Analytic Framework\label{sec:framework}}

The process studied in this paper can be illustrated as
\bea
p(P)+e(k)\rightarrow\eta_c(p)+X+e(k'),
\eea
which can be accessed through the evaluation of the following process,
\bea
p+\gamma^\star(q)\rightarrow\eta_c+X,
\eea
where $p$, $e$, and $\gamma^\star$ denote the proton, electron, and off-shell photon, respectively,
and $X$ can be any hadronic final states.
The effects of the initial protons can be further factorized as partonic processes convolved with the parton-distribution functions,
while the $\eta_c$ production can be factorized, according to nonrelativistic QCD~\cite{Bodwin:1994jh},
as the production of the intermediate $c\bar{c}$ states (the short-distance coefficients)
multiplied by the process-independent nonperturbative parameters (the long-distance matrix elements).
The factorization can be illustrated as
\bea
\mathrm{d}\sigma(p+\gamma^\star\rightarrow\eta_c+X)=\sum_{i,n}\int\mathrm{d}xf_{i/p}(x,\mu_f)\mathrm{d}\hat{\sigma}(i+\gamma^\star\rightarrow c\bar{c}[n]+X)\langle\mathcal{O}^{\eta_c}(n)\rangle,
\eea
where $i$ runs over all possible species of partons, namely gluon, $u$, $d$, and $s$ quarks,
$n$ denotes the quantum number of the intermediate $c\bar{c}$ state,
and $f_{i/p}(x,\mu_f)$ is the parton-distribution function at the factorization scale $\mu_f$.
The momentum of the initial parton, $p_i$ can be evaluated as $p_i=xP$.
For the $\eta_c$ production, four intermediate $c\bar{c}$ states are involved up to the order $\mathcal{O}(v^4)$,
where $v$ is the typical relative velocity of the charm quarks in the $\eta_c$ mesons,
they are $c\bar{c}[^1S_0^{[1]}]$, $c\bar{c}[^1S_0^{[8]}]$, $c\bar{c}[^3S_1^{[8]}]$, and $c\bar{c}[^1P_1^{[8]}]$.

The leading-order process for the $c\bar{c}[^1S_0^{[8]}]$ production is
\bea
g+\gamma^\star\rightarrow c\bar{c}[^1S_0^{[8]}],
\eea
which is of order $\mathcal{O}(\alpha\alpha_s)$.
Here, $g$ denotes a gluon, and $\alpha$ and $\alpha_s$ are the electroweak and strong couplings, respectively.
If we impose a cut-off on the transverse momentum of $\eta_c$ ($p_t^\star$), this process will not be observed.
Here and in the following, we use the superscript $\star$ to denote the kinematic variables in the proton-photon centre-of-mass frame.
The leading order processes for the production of the colour-octet intermediate states with nonzero $p_t^\star$ are listed below,
\bea
&&g+\gamma^\star\rightarrow c\bar{c}[^1S_0^{[8]}]+g, \NO \\
&&q+\gamma^\star\rightarrow c\bar{c}[^1S_0^{[8]}]+q, \NO \\
&&g+\gamma^\star\rightarrow c\bar{c}[^3S_1^{[8]}]+g, \NO \\
&&q+\gamma^\star\rightarrow c\bar{c}[^3S_1^{[8]}]+q, \NO \\
&&g+\gamma^\star\rightarrow c\bar{c}[^1P_1^{[8]}]+g,
\eea
where $q$ denotes a light quark ($u$, $d$, $s$).
The leading-order process to produce a colour-singlet $\eta_c$ is
\bea
g+\gamma^\star\rightarrow c\bar{c}[^1S_0^{[1]}]+g+g.
\eea

Denoting the momenta of the initial proton and electron, the short-distance coefficients for the processes listed above can be written as
\bea
\mathrm{d}\hat{\sigma}(i+\gamma^\star\rightarrow c\bar{c}[n]+X)=
\frac{1}{2xS}\frac{1}{N_sN_c}\frac{1}{(Q^2)^2}L_{\mu\nu}H^{\mu\nu}(i+\gamma^\star\rightarrow c\bar{c}[n]+X)\mathrm{d}\Phi,
\eea
where $S$ and $Q^2$ are defined in terms of the following equation,
\bea
S=(P+k)^2\approx2P\cdot k,~~~~Q^2=(k-k')^2,
\eea
$1/(N_sN_c)$ is the colour and spin average factor,
$L_{\mu\nu}$ and $H^{\mu\nu}$ are the leptonic and hadronic tensors, respectively,
and $\mathrm{d}\Phi$ is the phase space.

Having the definition of the following invariants,
\bea
W^2=(P+q)^2,~~~~y=\frac{W^2+Q^2}{S},~~~~z=\frac{P\cdot p}{P\cdot q},
\eea
the leptonic tensor can be expressed as
\bea
L_{\mu\nu}=8\pi\alpha Q^2l_{\mu\nu},
\eea
with
\bea
l^{\mu\nu}=A_g(-g^{\mu\nu}-\frac{q^\mu q^\nu}{Q^2})+A_l\epsilon_L^\mu\epsilon_L^\nu+A_t\epsilon_T^\mu\epsilon_T^\nu+A_{lt}(\epsilon_l^\mu\epsilon_t^\nu+\epsilon_t^\mu\epsilon_l^\nu).
\eea
where
\bea
&&\epsilon_l=\frac{1}{Q}(q+\frac{2Q^2}{s}p_i), \NO \\
&&\epsilon_t=\frac{1}{p_t^\star}(p-\rho p_i-zq),
\eea
and
\bea
&&A_g=1+\frac{2(1-y)}{y^2}-\frac{2(1-y)}{y^2}\cos(2\psi^{\star}), \NO \\
&&A_L=1+\frac{6(1-y)}{y^2}-\frac{2(1-y)}{y^2}\cos(2\psi^{\star}), \NO \\
&&A_{LT}=\frac{2(2-y)}{y^2}\sqrt{1-y}\cos(\psi^{\star}), \NO \\
&&A_T=\frac{4(1-y)}{y^2}\cos(2\psi^{\star}).
\eea
Here, $\psi^\star$ is the azimuthal angle of $\eta_c$ in the proton-photon centre-of-mass frame, and
\bea
&&s=(p_i+q)^2+Q^2=2p_i\cdot q, \NO \\
&&m_t^{\star2}=p_t^{\star2}+M^2 \NO \\
&&\rho=\frac{m_t^{\star2}/z+zQ^2}{s},
\eea
where $M$ is the $\eta_c$ mass.

For the colour-octet processes~\cite{Sun:2017nly},
\bea
\mathrm{d}xf_{i/p}(x)\mathrm{d}\Phi=\frac{1}{(4\pi)^4S(W^2+Q^2)z(1-z)}f_{i/p}(x)\mathrm{d}Q^2\mathrm{d}W^2\mathrm{d}z\mathrm{d}p_t^{\star2}\mathrm{d}\psi^\star,
\eea
while for the colour-singlet process, i.e.
\bea
g(p_i)+\gamma^\star(q)\rightarrow c\bar{c}[^1S_0^{[1]}](p)+g(p_1)+g(p_2),
\eea
we have
\bea
\mathrm{d}xf_{i/p}(x,\mu_f)\mathrm{d}\Phi=\frac{1}{(4\pi)^7S(W^2+Q^2)z(1-z)}f_{i/p}(x)\mathrm{d}Q^2\mathrm{d}W^2\mathrm{d}z\mathrm{d}p_t^{\star2}\mathrm{d}\psi^\star\mathrm{d}s_{12}\mathrm{d}\Omega_1,
\eea
where $s_{12}=(p_1+p_2)^2$,
and $\mathrm{d}\Omega_{1}$ is the infinitesimal spatial angle of $\bm{p}_1$ in the $p_1$-$p_2$ centre-of-mass frame.

The results of the hadronic tensors for $c\bar{c}[^1S_0^{[8]}]$ and $c\bar{c}[^3S_1^{[8]}]$ have been given in Reference~\cite{Sun:2017nly}.
The only missing elements are those for $c\bar{c}[^1S_0^{[1]}]$ and $c\bar{c}[^1P_1^{[8]}]$.
Since the results for the colour-singlet state is very complicated,
we do not present their analytical results.
These missing elements are calculated using a new \textit{Mathematica} program.
Its validity is checked by the reevaluation of the processes for the production of the $^1S_0^{[8]}$ and $^3S_1^{[8]}$ states.

\section{Numerical Results\label{sec:results}}

In our numerical calculations, we adopt the following parameter choices.
The default value of the charm quark mass is given as $m_c=1.5\gev$,
and the fine-structure constant is approximated as $\alpha=1/137$.
The renormalization scale ($\mu_r$) and factorization scale ($\mu_f$) are set to be $\mu_r=\mu_f=\mu_0\equiv\sqrt{Q^2+M^2}$, where $M$ is the $\eta_c$ mass.
For the sake of gauge invariance, its value is fixed to $M=2m_c$.
For HERA experiment, the energy of the electron beams is $27.5\gev$ and that of the proton beams is $920\gev$,
while for the EIC, they are $21\gev$ and $100\gev$, respectively.
We employ CTEQ6L1~\cite{Pumplin:2002vw} as the PDF for the protons.
The CS LDME is computed according to
\bea
\langle\mathcal{O}^{\eta_c}(^1S_0^{[1]})\rangle=\frac{3}{2\pi}\big|R(0)\big|^2,
\eea
To obtain the value of $\alpha_s$, one-loop running equation is employed and its value at the $Z_0$-boson mass is set to be $\alpha_s(M_Z)=0.13$.
For the HERA experiment, we apply the following kinematic constraints,
$4\gev^2<p_t^{\star2}<100\gev^2$, $4\gev^2<Q^2<100\gev^2$, $60\gev<W<240\gev$, and $0<z<0.6$, respectively,
while for the EIC experiment, the constraints are $4\gev^2<p_t^{\star2}<36\gev^2$, $4\gev^2<Q^2<36\gev^2$, $20\gev<W<80\gev$, and $0<z<0.6$, respectively.
In this paper, we calculate the differential cross sections with respect to $p_t^2$, $p_t^{\star2}$, $Q^2$, $W$, and $z$,
in the kinematic regions constrained by the conditions given above, excluding that for the observed variable.
For instance, the differential cross section with respect to $p_t^{\star2}$ is calculated in the region defined by the following constraints,
$4\gev^2<Q^2<36\gev^2$, $20\gev<W<80\gev$, and $0<z<0.6$.
Note that we use $p_t$ to denote the transverse momentum of $\eta_c$ measured in the laboratory frame.

\subsection{Results for HERA experiment}

\begin{figure}
\includegraphics[scale=0.5]{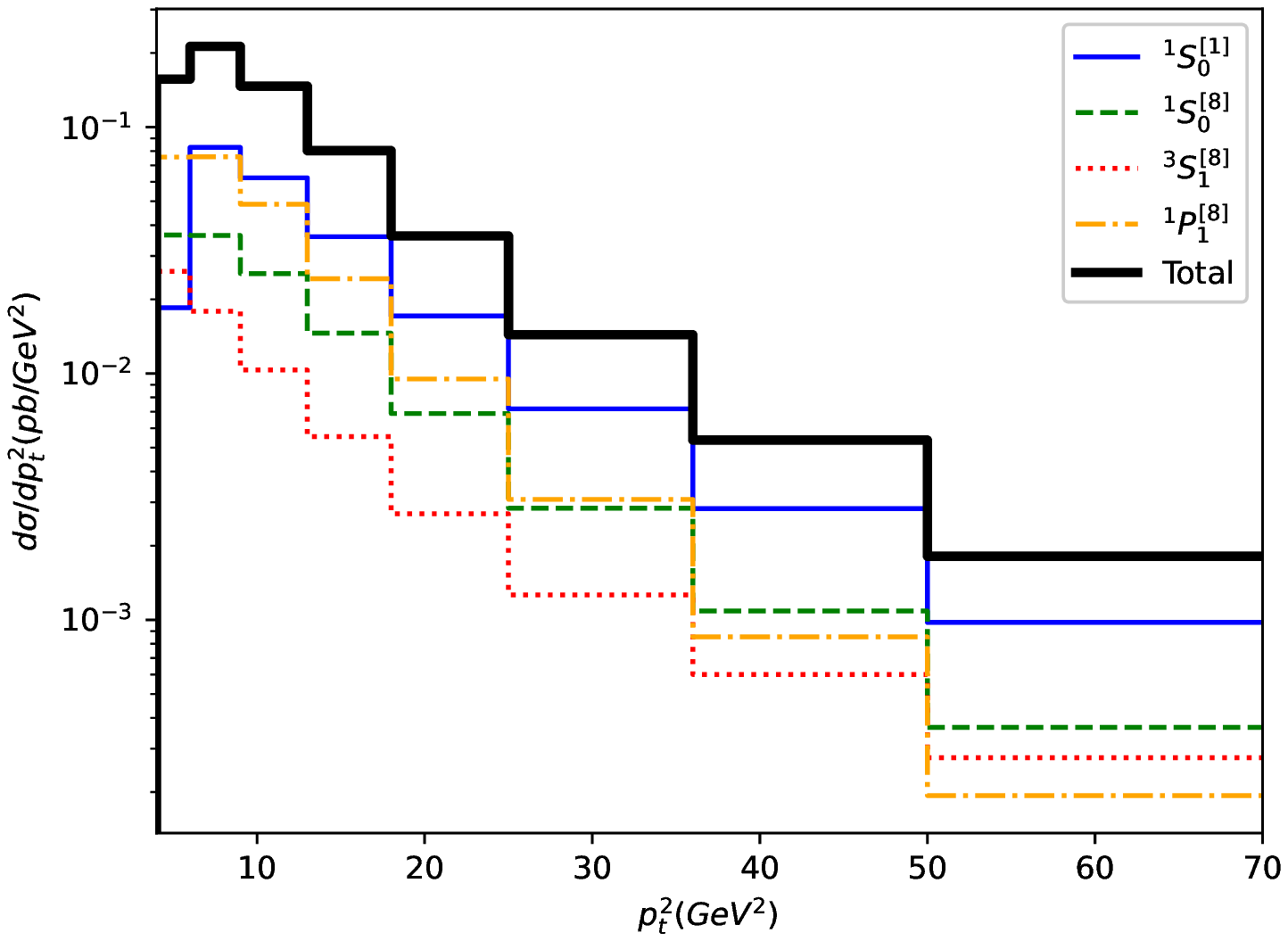}
\includegraphics[scale=0.5]{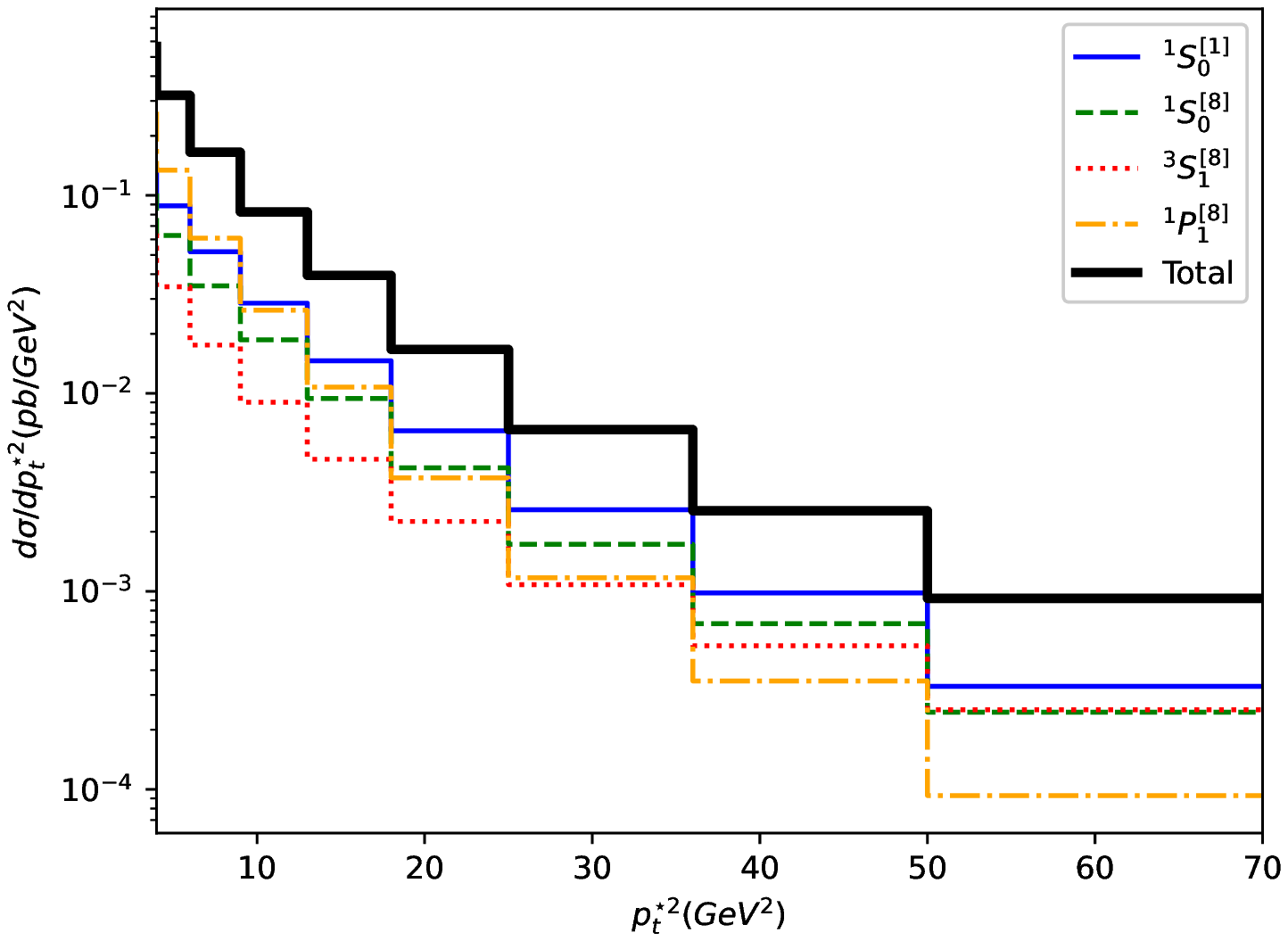}\\
\includegraphics[scale=0.5]{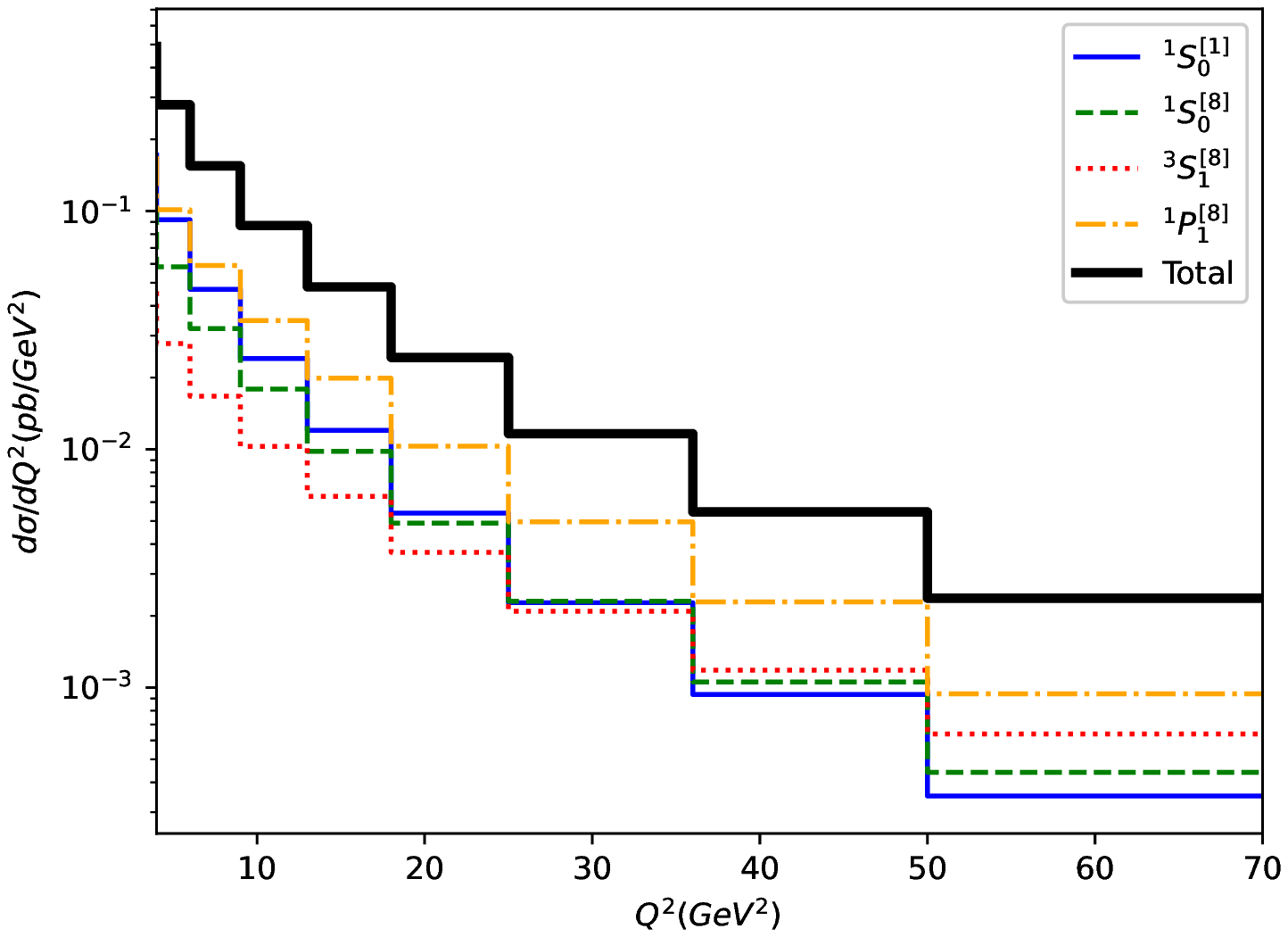}
\includegraphics[scale=0.5]{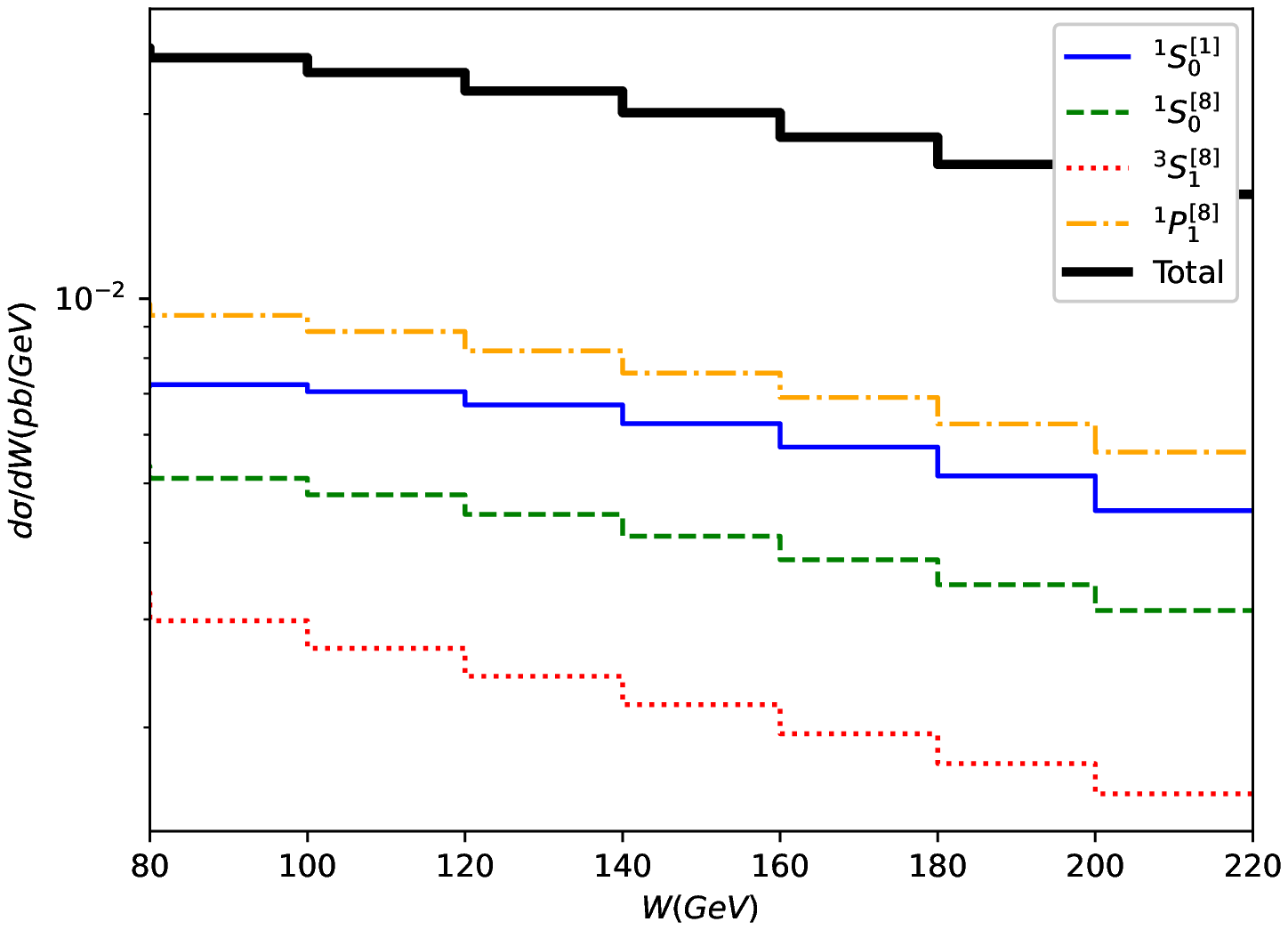}\\
\includegraphics[scale=0.5]{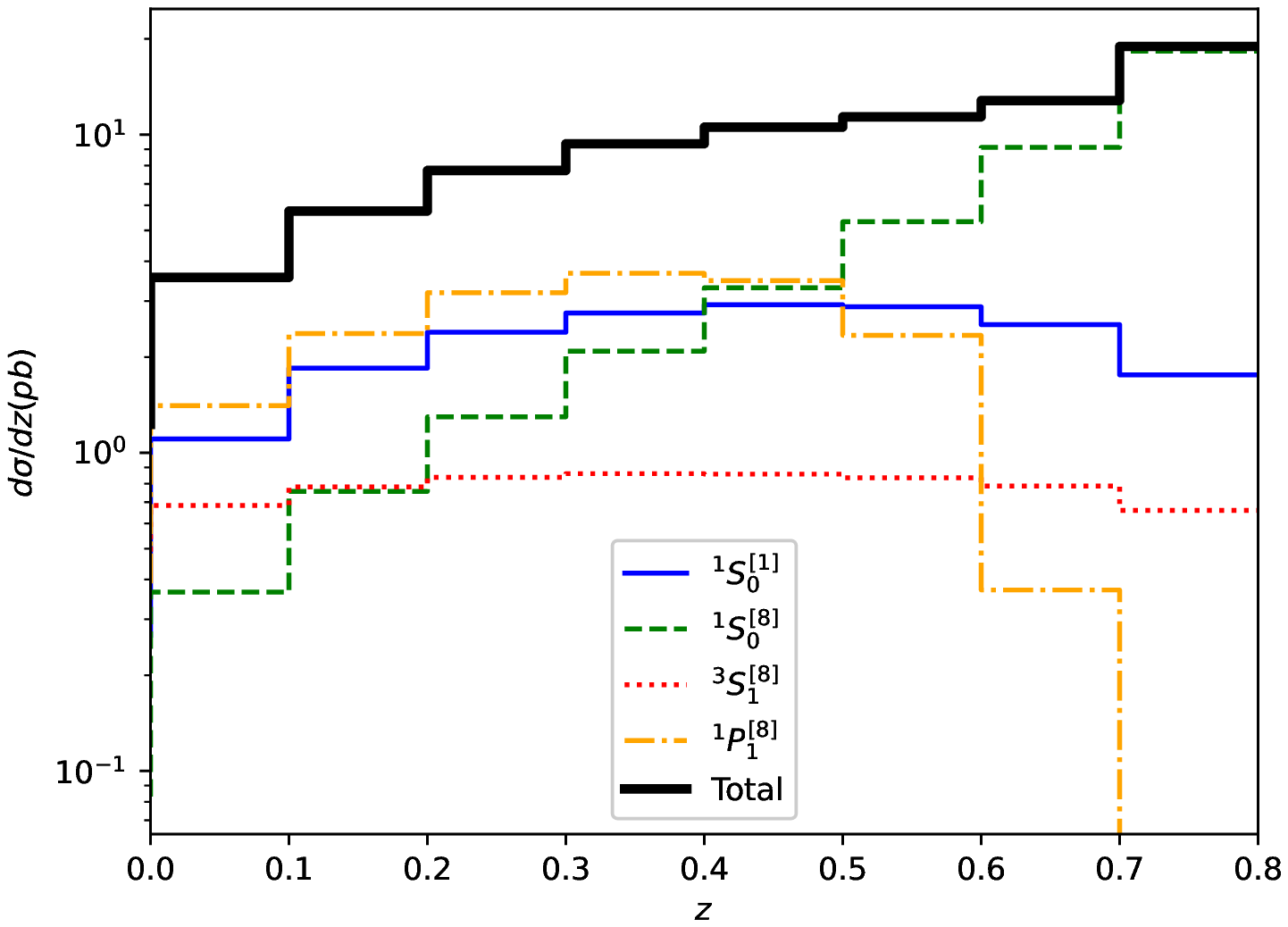}
\caption{\label{fig:HERA}
The differential cross sections of the $\eta_c$ leptoproduction with respect to $p_t^2$,
$p_t^{\star2}$, $Q^2$, $W$, and $z$ in HERA experimental condition.
}
\end{figure}
The results for the HERA experimental condition are presented in Figure~\ref{fig:HERA},
where the long-distance matrix elements are taken from References~\cite{Zhang:2014ybe, Sun:2015pia}.
Their values are
\bea
&&\langle\mathcal{O}^{\eta_c}(^1S_0^{[1]})\rangle=0.16\gev^3, \NO \\
&&\langle\mathcal{O}^{\eta_c}(^1S_0^{[8]})\rangle=0.36\times10^{-2}\gev^3, \NO \\
&&\langle\mathcal{O}^{\eta_c}(^3S_1^{[8]})\rangle=0.74\times10^{-2}\gev^3, \NO \\
&&\langle\mathcal{O}^{\eta_c}(^1P_1^{[8]})\rangle/m_c^2=6.0\times10^{-2}\gev^3. \label{eqn:LDMEz}
\eea

In low $p_t$ ($p_t^\star$) regions, the most important contribution comes from the intermediate state $c\bar{c}[^1P_1^{[8]}]$,
which is mainly due to a large long-distance matrix element.
As $p_t$ ($p_t^\star$) increases, its importance rapidly decreases,
and the colour-singlet part, although enduring a small long-distance matrix element comparing to the values obtained via potential model,
takes the superiority over all its colour-octet counterparts.
This observation is unexpected.
People used to consider the colour-singlet contribution to be negligible in this process (see e.g. \cite{Hao:2000ci}),
which is true in a similar process, the $\eta_c$ production in $ep$ collisions via direct photons~\cite{Zhang:2019wxo}.
Although seemingly weird, it is not difficult to understand.
As a matter of fact, a term scaling as $p_t^{-6}$ ($p_t^{\star-6}$) arises from the off-shell photon in the colour-singlet $\eta_c$ production,
and vanishes as $Q^2\to0$.
This explains why in the $\eta_c$ photoproduction the colour-singlet differential cross section with respect to $p_t$ suffers a sharper decrease than the $^1S_0^{[8]}$ channel,
while in the $\eta_c$ leptoproduction, these two channels has almost the same $p_t$ ($p_t^\star$) behaviour.

One might notice that when $z$ approaches 1, the cross section for the $^1S_0^{[8]}$ channel grows rapidly.
This behaviour is due to an unphysical divergence at the endpoint $z=1$,
and will be greatly suppressed if this divergence is smeared by employing some shape functions.
For this reason, we impose a cutoff, $z<0.6$, when evaluating the differential cross sections with respect to all the other kinematic variables.

Another interesting feature that can be seen in Figure~\ref{fig:HERA} is that each of the four channels has a unique behaviour.
For example, the differential cross section with respect to $p_t^2$ for the $^1S_0^{[1]}$,
$^1S_0^{[8]}$, $^3S_1^{[8]}$, and $^1P_1^{[8]}$ channels scale as $p_t^{-6}$, $p_t^{-6}$, $p_t^{-4}$, and $p_t^{-8}$, respectively, in hight $p_t$ limit;
the colour-singlet and colour-octet $^1S_0$ states can be further distinguished in the $Q^2$ distributions.
This feature provides us a very good laboratory to study the long-distance matrix elements for the S-wave charmonia production.

\subsection{Results for EIC experiment}

\begin{figure}
\includegraphics[scale=0.5]{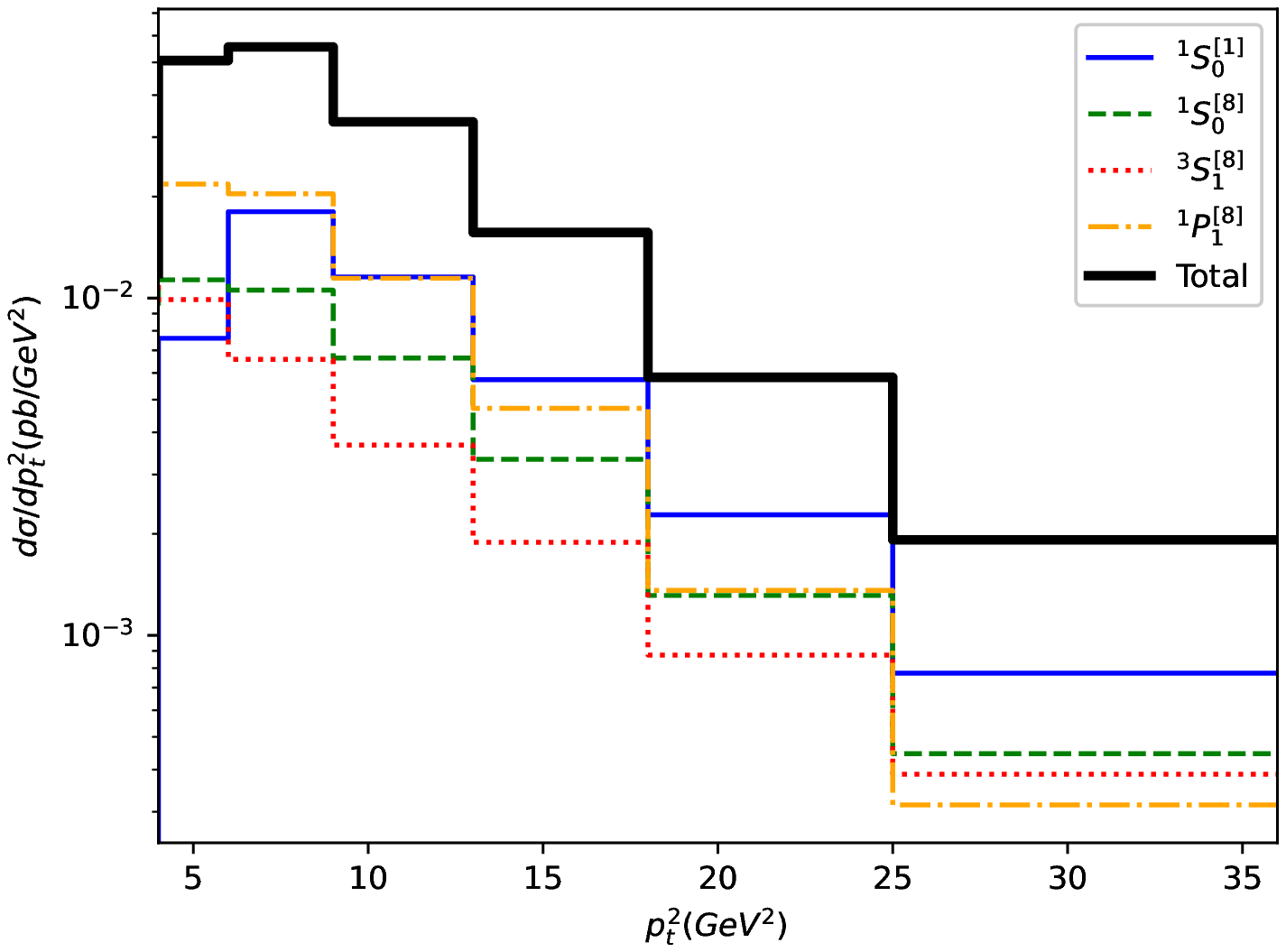}
\includegraphics[scale=0.5]{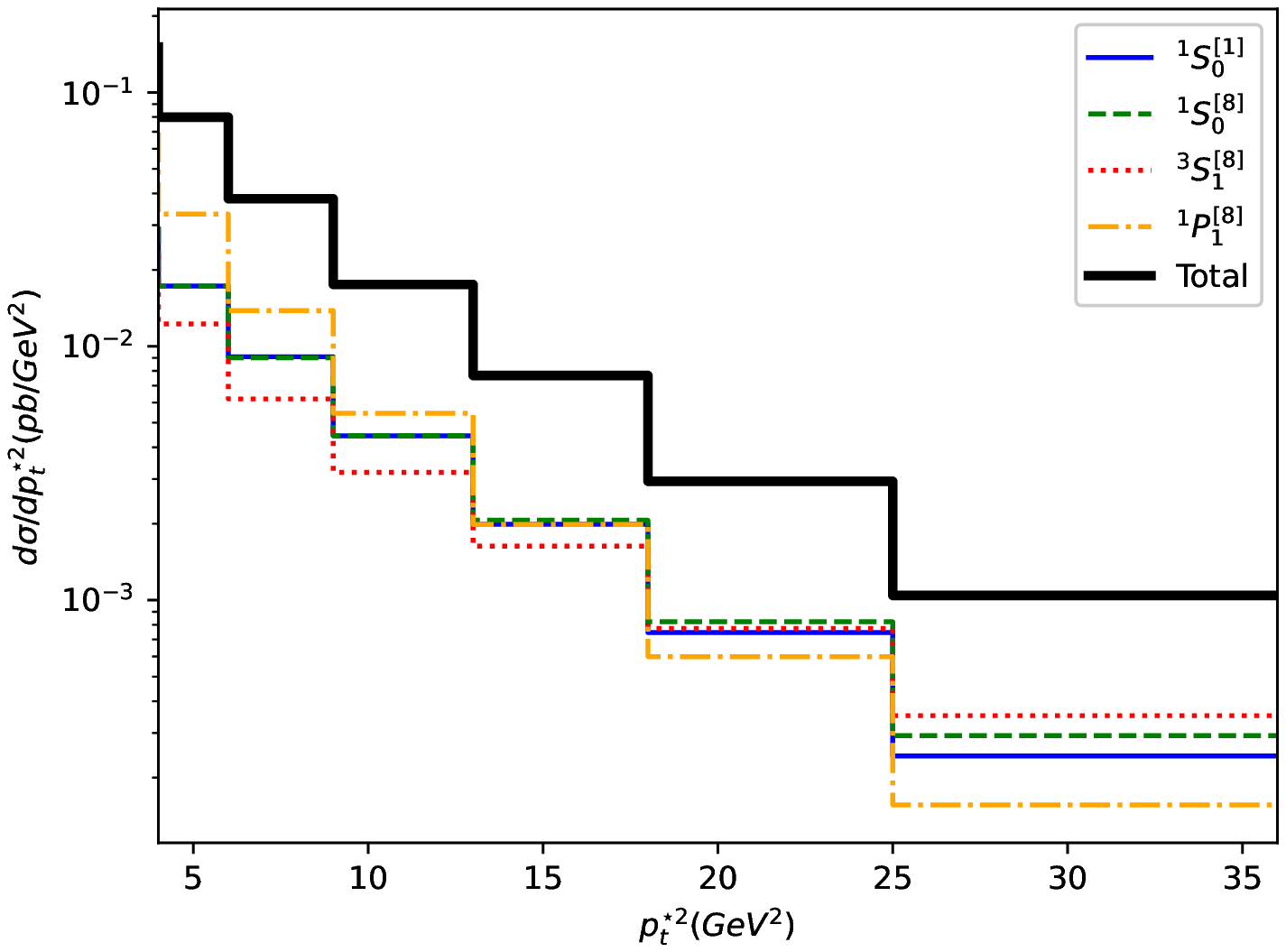}\\
\includegraphics[scale=0.5]{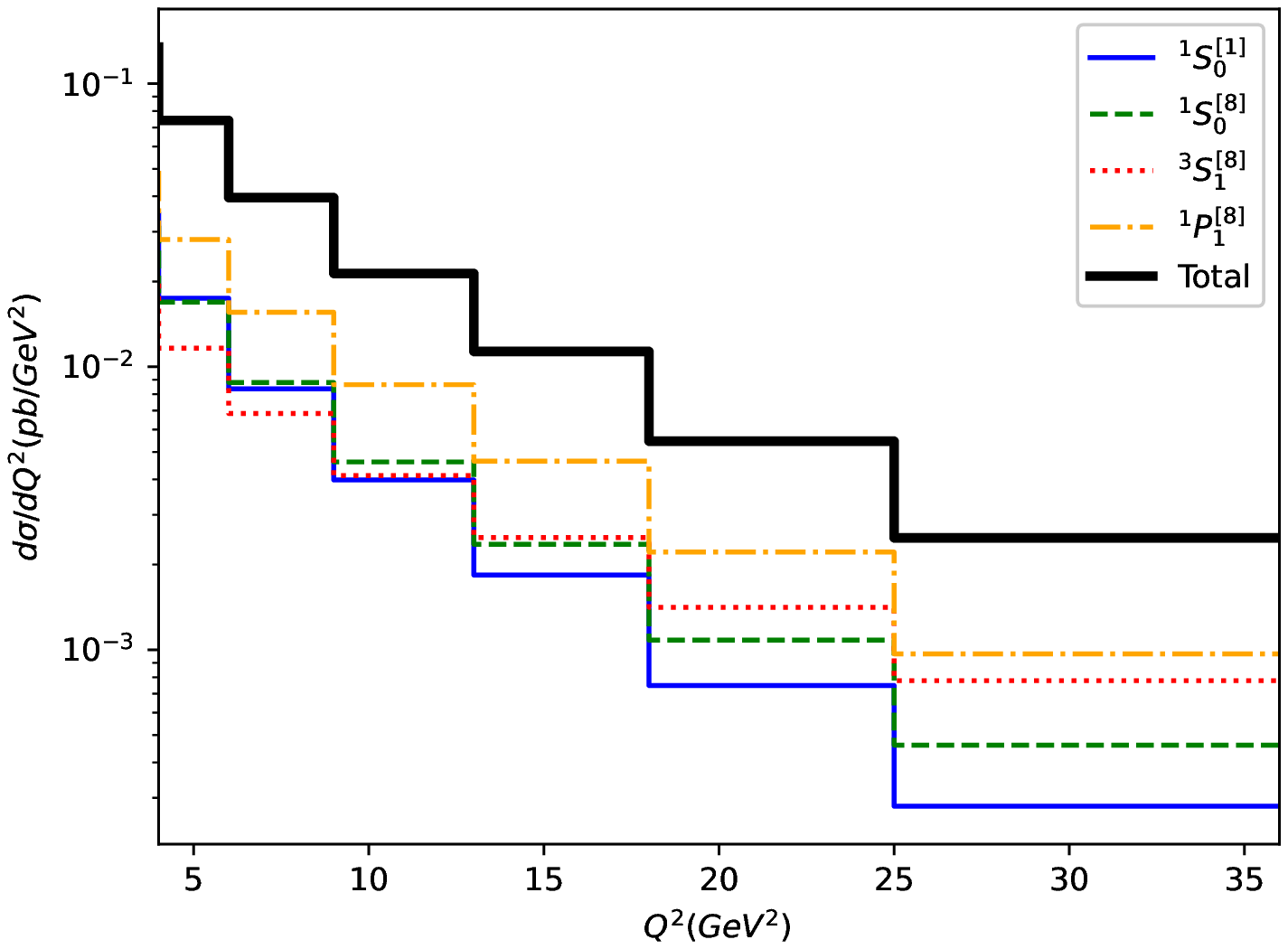}
\includegraphics[scale=0.5]{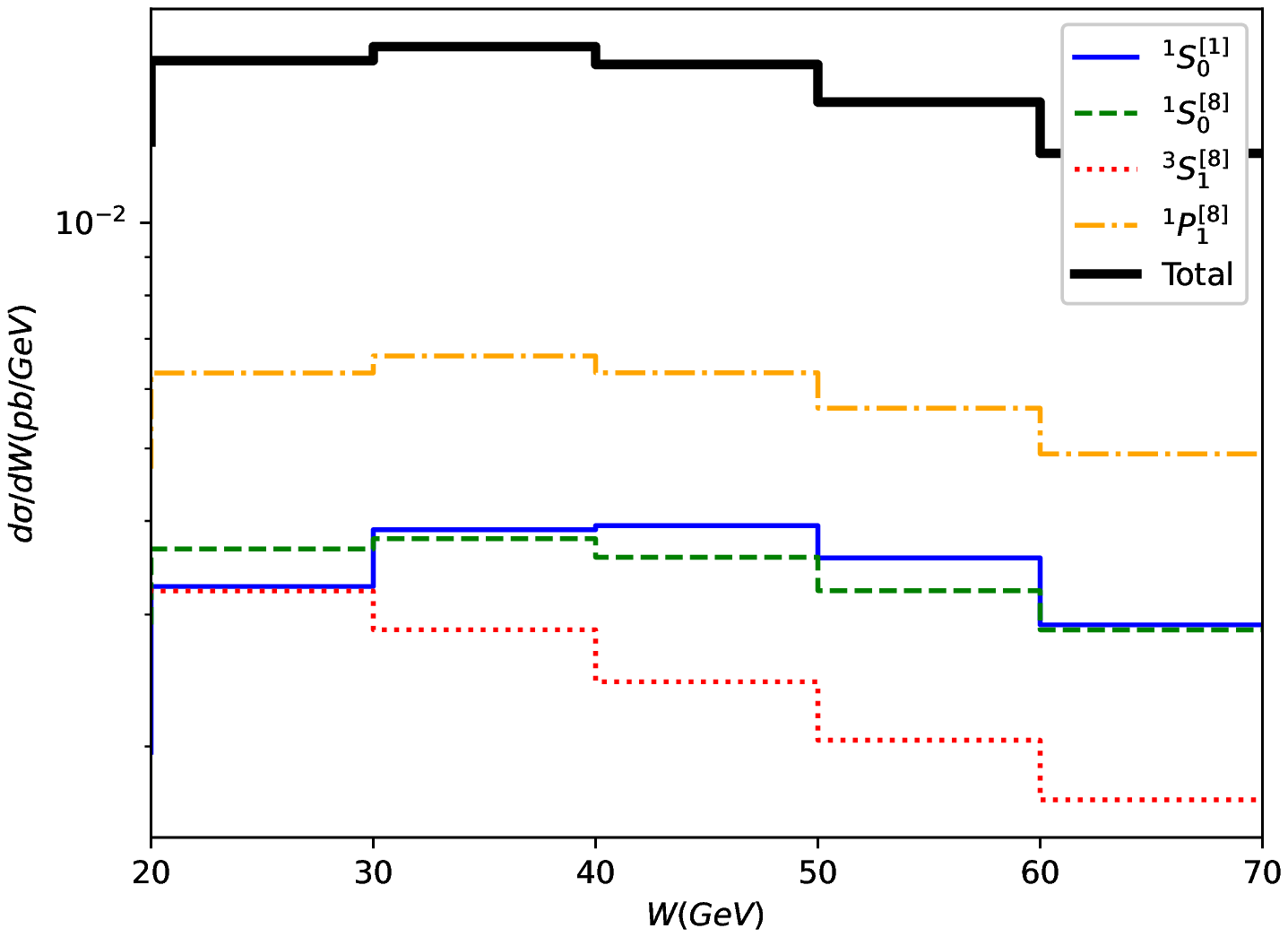}\\
\includegraphics[scale=0.5]{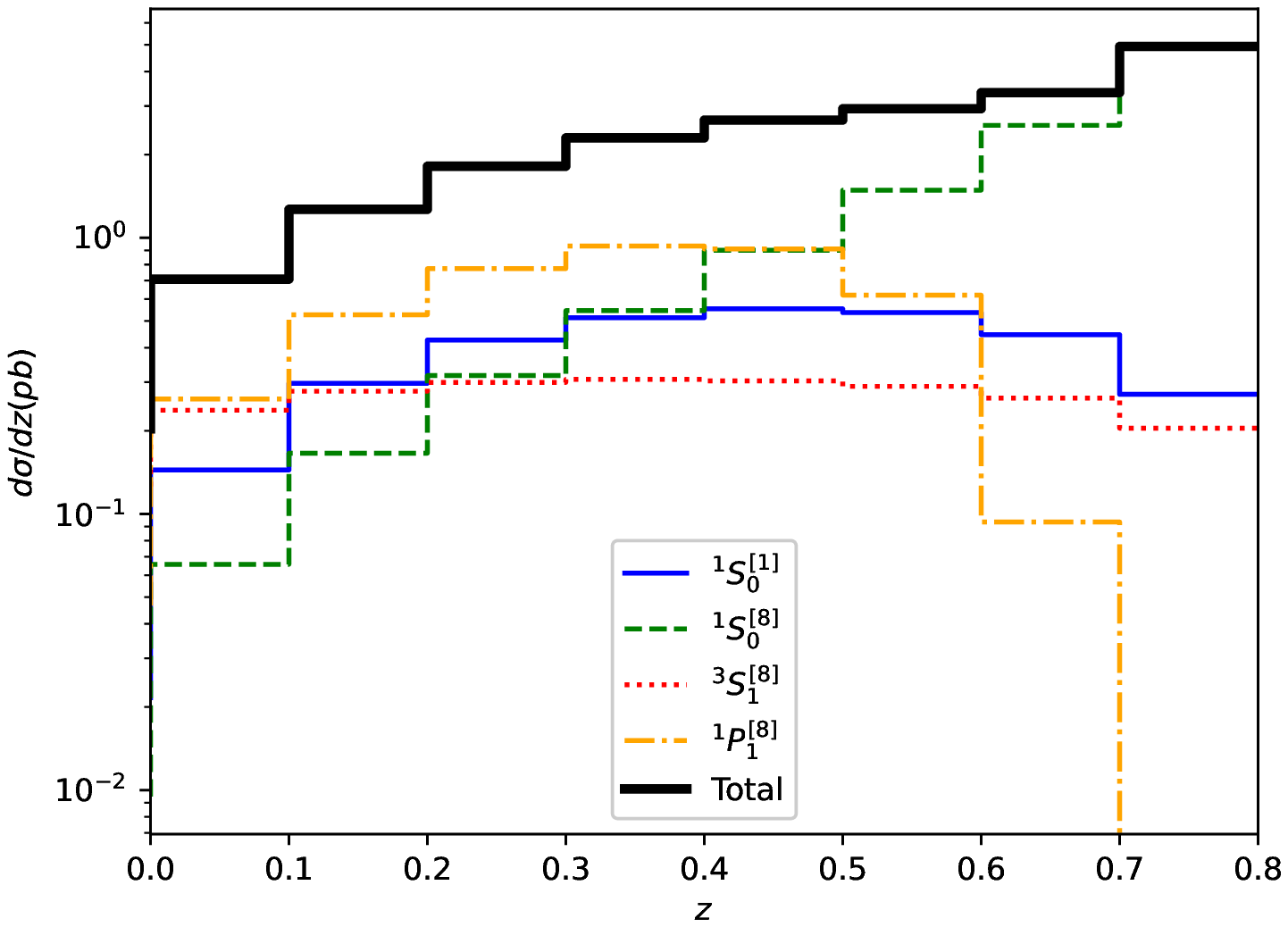}
\caption{\label{fig:EIC}
The differential cross sections of the $\eta_c$ leptoproduction with respect to $p_t^2$,
$p_t^{\star2}$, $Q^2$, $W$, and $z$ in EIC experimental condition.
}
\end{figure}
In the EIC experimental condition (Figure~\ref{fig:EIC}), the cross sections are slightly smaller than the HERA ones,
keep all the relative significances of the four channels.
Since the EIC will run in much higher luminosity than the HERA experiment,
it is promising to expect a high-accuracy measurement of the $\eta_c$ leptoproduction.

\subsection{Scale dependences}

\begin{figure}
\includegraphics[scale=0.5]{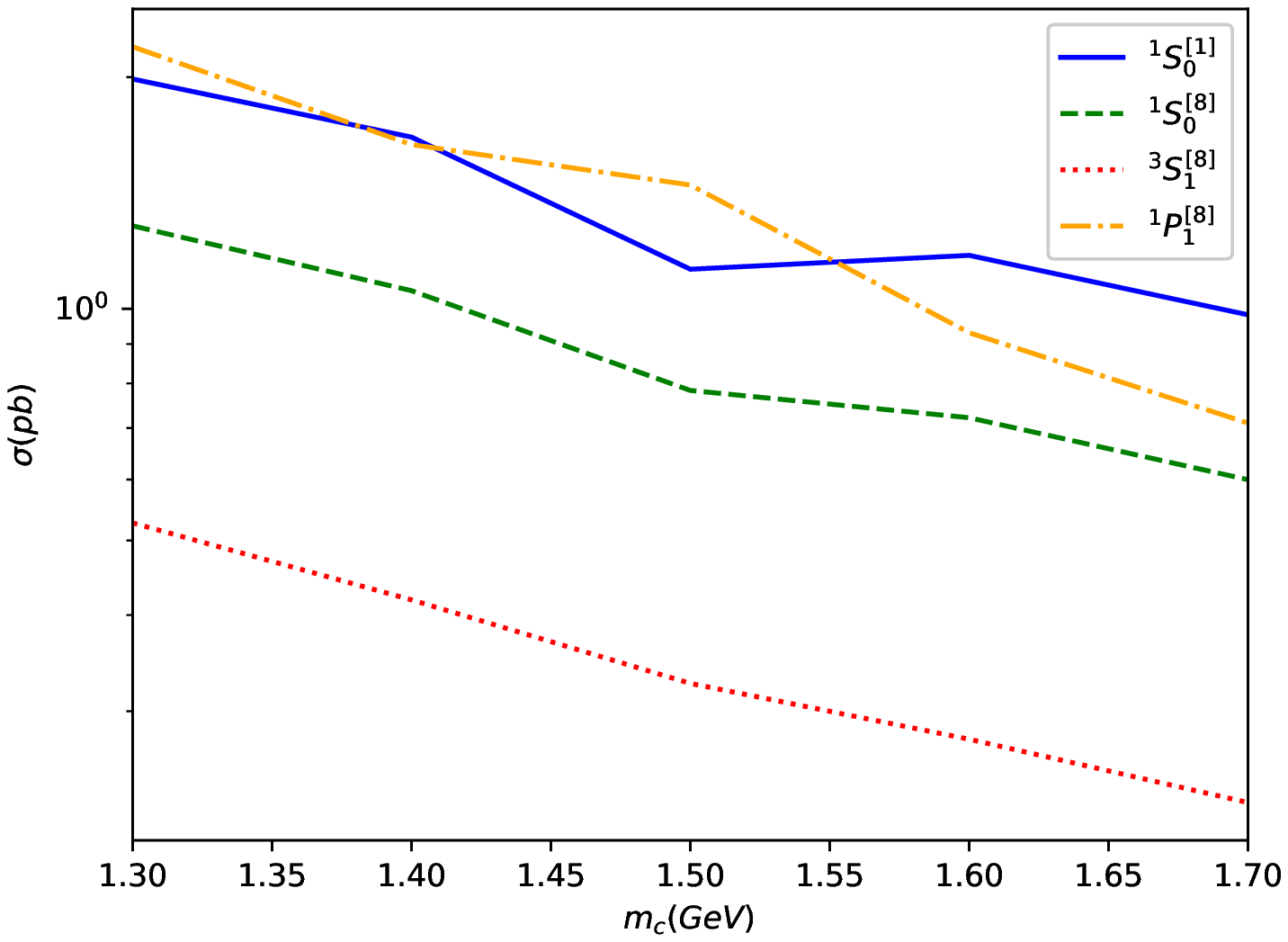}
\includegraphics[scale=0.5]{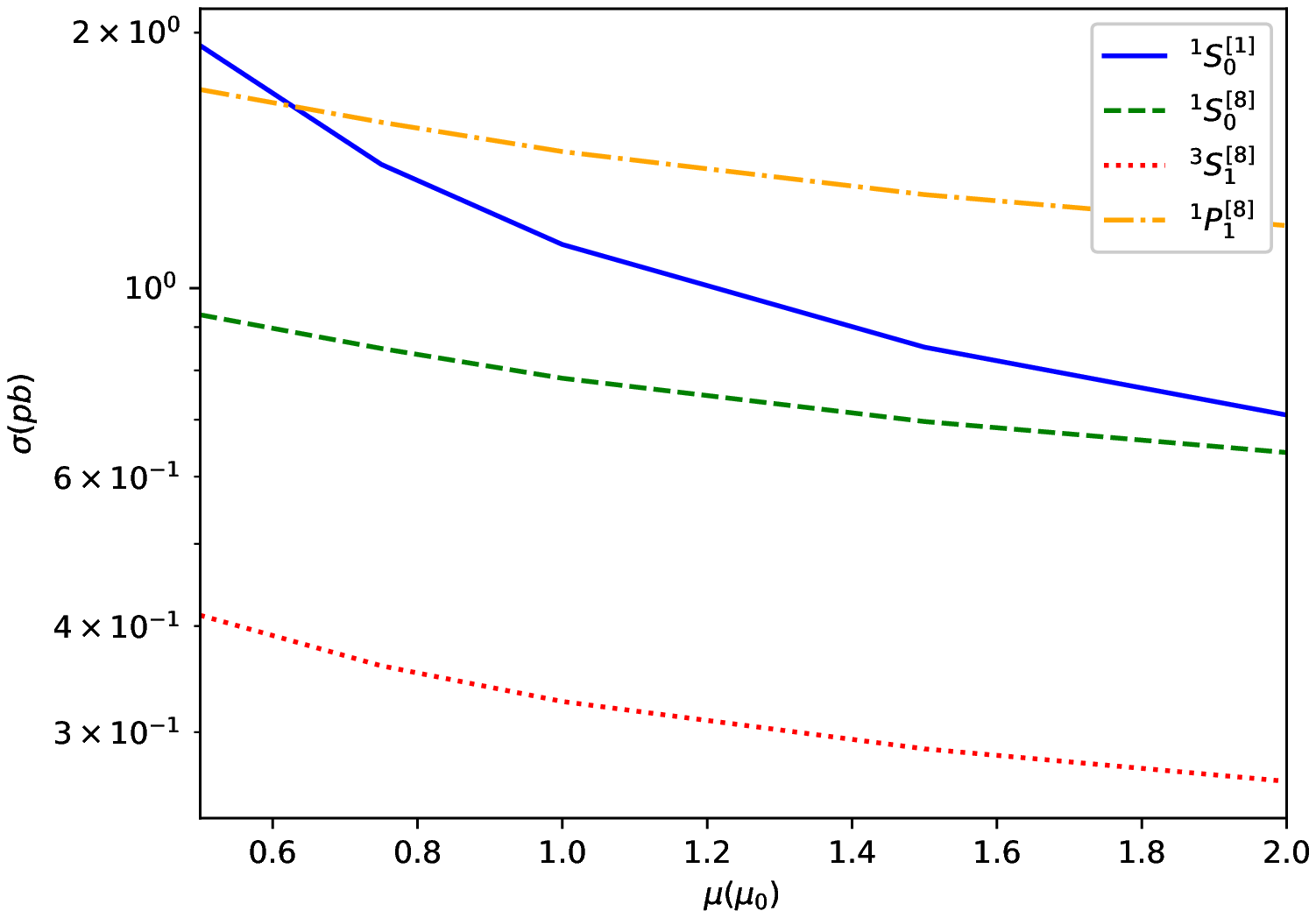}
\caption{\label{fig:scaledep}
The dependence of the integrated cross section on $m_c$ and $\mu_r=\mu_f=\mu$.
}
\end{figure}
We also calculated the integrated cross sections, in the HERA experimental condition,
for the four intermediate states with different choices of the of the values of the charm quark mass and the renormalisation and factrorisation scales.
$m_c$ varies from $1.3\gev$ to $1.7\gev$, and $\mu_r=\mu_f=\mu$ varies from $0.5\mu_0$ to $2\mu_0$, where $\mu_0=\sqrt{Q^2+M^2}$.
The kinematic constraints of the integral follows the conditions given above,
namely, $4\gev^2<p_t^{\star2}<100\gev^2$, $4\gev^2<Q^2<100\gev^2$, $60\gev<W<240\gev$, and $0<z<0.6$.
It is found that the scale dependence of the $\eta_c$ leptoproduction is mild,
which indicate good convergence of the perturbative expansions.

\subsection{Long-distance-matrix-element dependence}

Since there are several sets of long-distance matrix elements available on the market,
it is necessary to compare their corresponding predictions and see whether,
just like in the $J/\psi$ hadroproduction case, they lead to almost the same results.
In the following discussions, we employ five different sets of long-distance matrix elements,
taken from References~\cite{Butenschoen:2011yh, Chao:2012iv, Bodwin:2015iua, Feng:2018ukp}, respectively.
They are listed in Table~\ref{tab:ldme}, excluding our default set which has been given in Equation~\ref{eqn:LDMEz}.
Each of them is independently obtained, and can describe the $J/\psi$ yield data at the Tevatron and LHC.
Note that the long-distance matrix elements for the $J/\psi$ production are extracted also in References~\cite{Gong:2012ug, Bodwin:2014gia},
since the authors of the two references have updated their results in References~\cite{Feng:2018ukp, Bodwin:2015iua}, respectively,
we do not present results for the old version of the long-distance matrix elements in this paper.

\begin{table}
\begin{center}
\caption{
\label{tab:ldme}
The colour-octet long-distance matrix elements for the $\eta_c$ production obtained by different theory groups,
where the relations between the LDMEs for the $J/\psi$ and $\eta_c$ production are employed.
}
\begin{tabular}{cccccc}
\hline
\hline
References&~Butenschon~&~Chao and~&Bodwin and~&Feng and \\
~&and \textit{et al.}~\cite{Butenschoen:2011yh}&~\textit{et al.}~\cite{Chao:2012iv}~
&~\textit{et al.}~\cite{Bodwin:2015iua}~&~\textit{et al.}~\cite{Feng:2018ukp} \\
\hline
$\langle{\cal O}^{\eta_c}(^1S_0^{[1]})\rangle(\gev^3)$~&~0.44~&~0.387~&~0.387~&~0.387 \\
$\langle{\cal O}^{\eta_c}(^1S_0^{[8]})\rangle(10^{-2}\gev^3)$~&~$0.056\pm0.015$~&~$0.10\pm0.04$~&~$-0.238\pm0.121$~&~$0.059\pm0.019$ \\
$\langle{\cal O}^{\eta_c}(^3S_1^{[8]})\rangle(10^{-2}\gev^3)$~&~$3.04\pm0.35$~&~$8.9\pm0.98$&~$11.0\pm1.4$~&~$5.66\pm0.47$ \\
$\langle{\cal O}^{\eta_c}(^1P_1^{[8]})\rangle/m_c^2(10^{-2}\gev^3)$~&~$-1.21\pm0.21$~&~$1.68\pm0.63$~&~$-0.936\pm0.453$~&~$1.03\pm0.31$ \\
\hline
\hline
\end{tabular}
\end{center}
\end{table}

\begin{figure}
\includegraphics[scale=0.5]{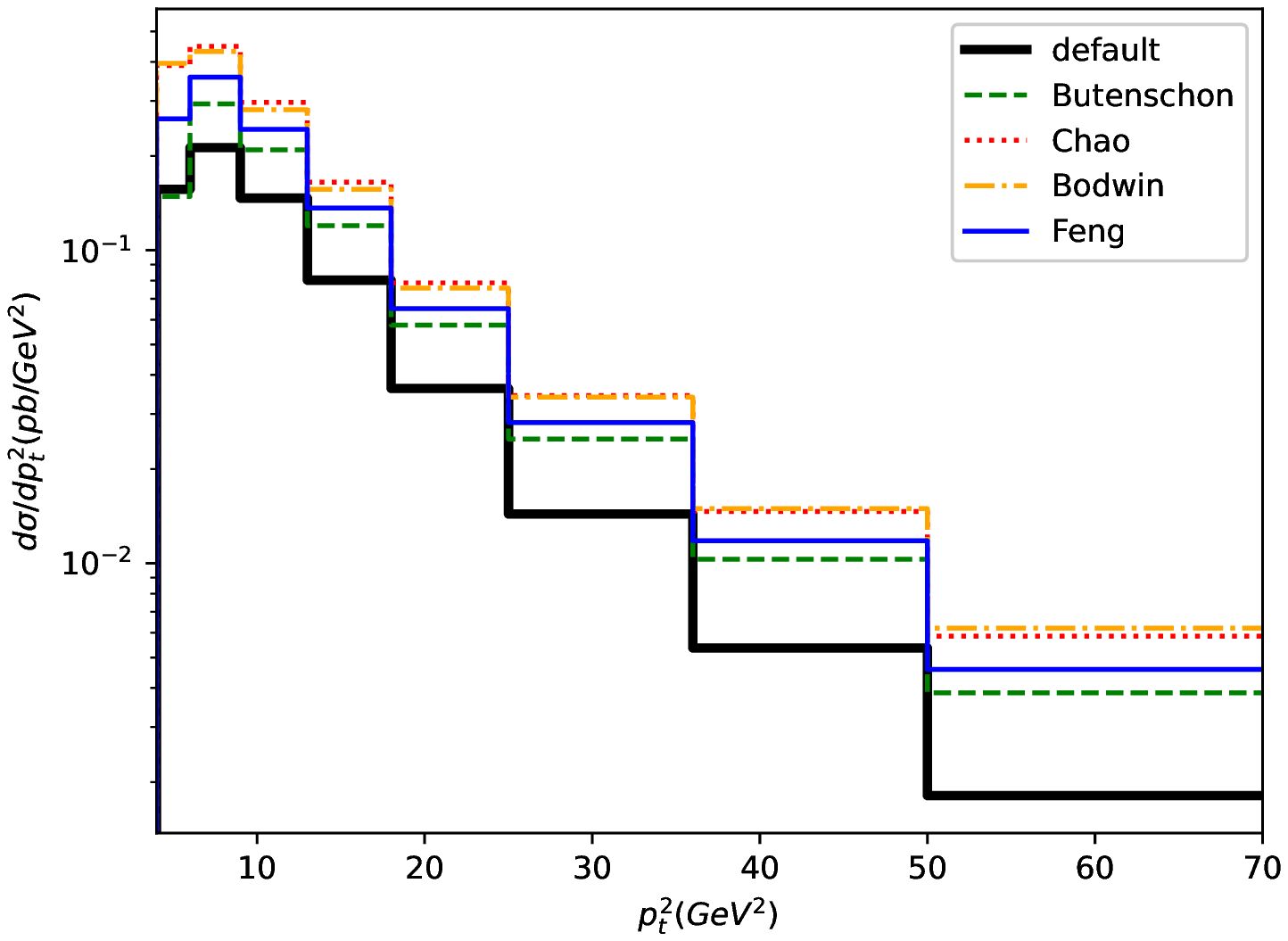}
\includegraphics[scale=0.5]{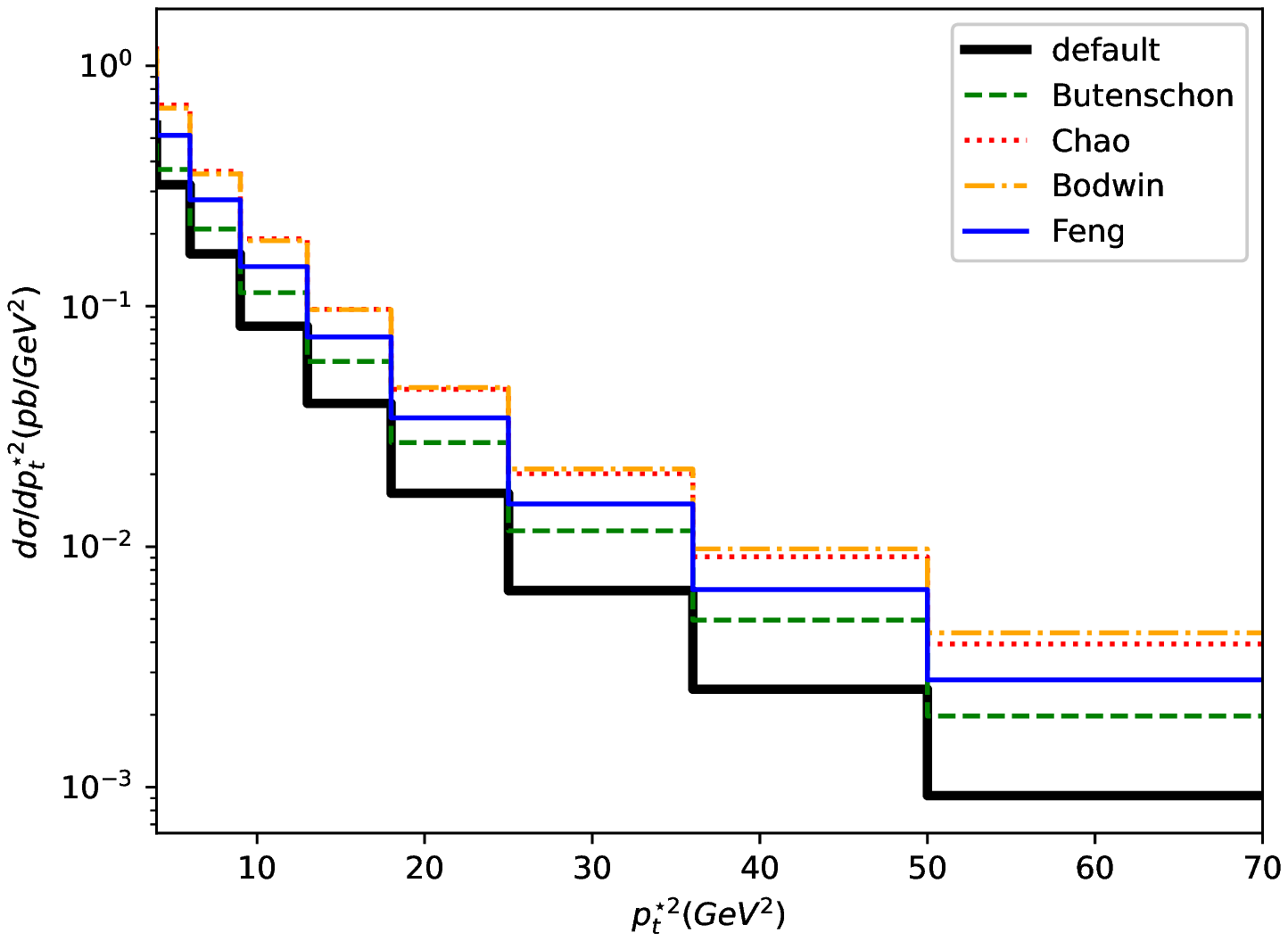}\\
\includegraphics[scale=0.5]{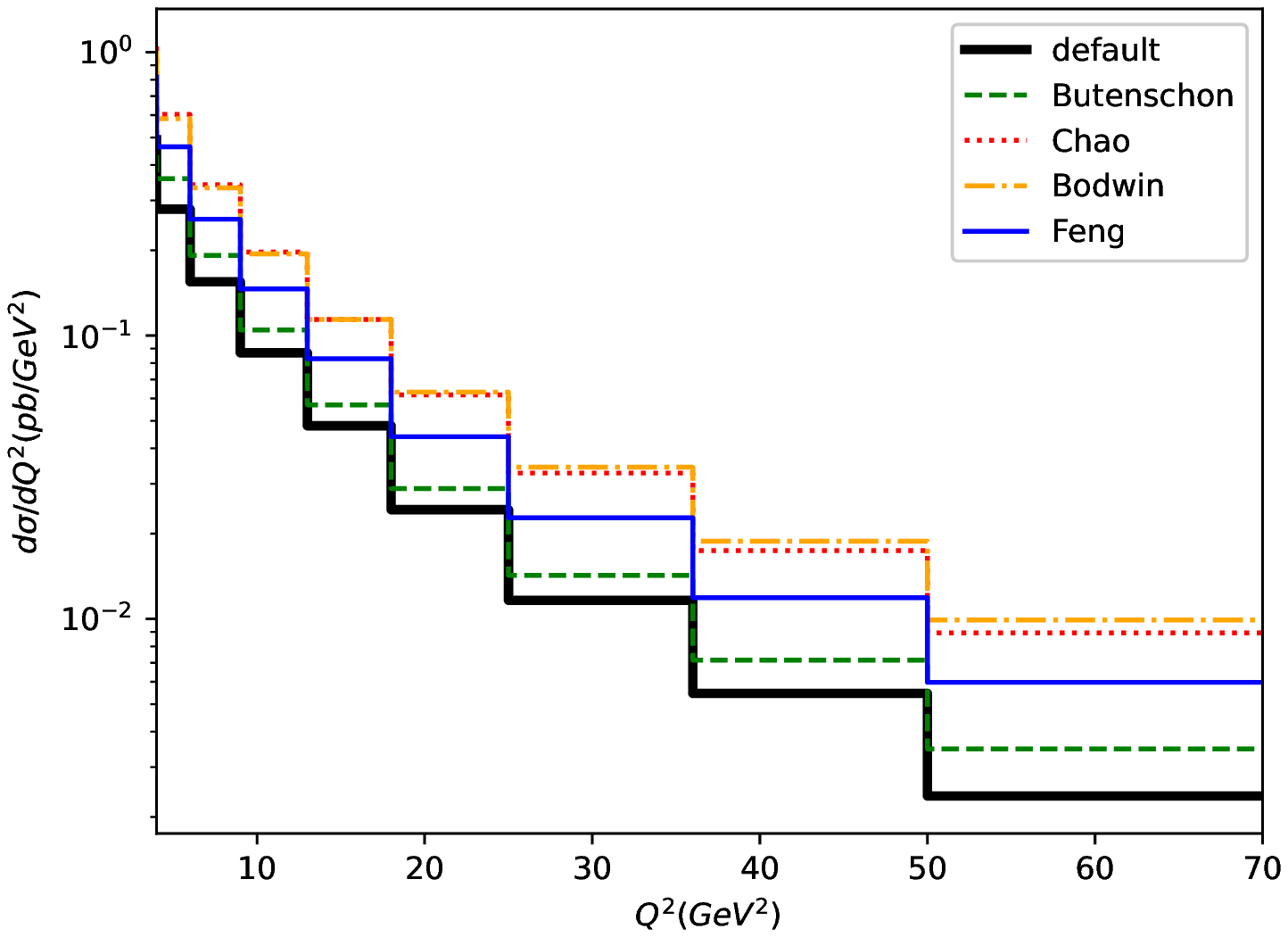}
\includegraphics[scale=0.5]{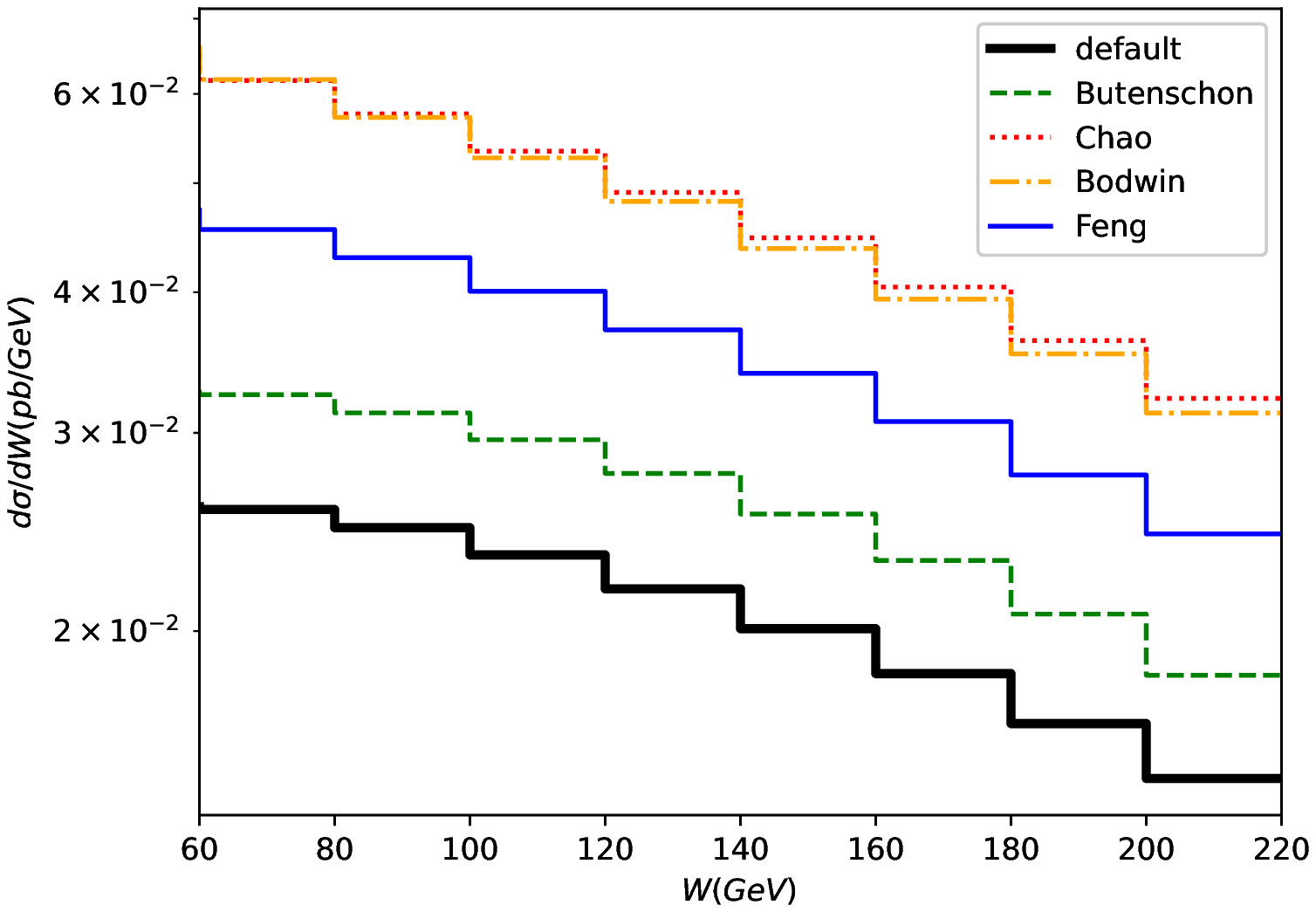}\\
\includegraphics[scale=0.5]{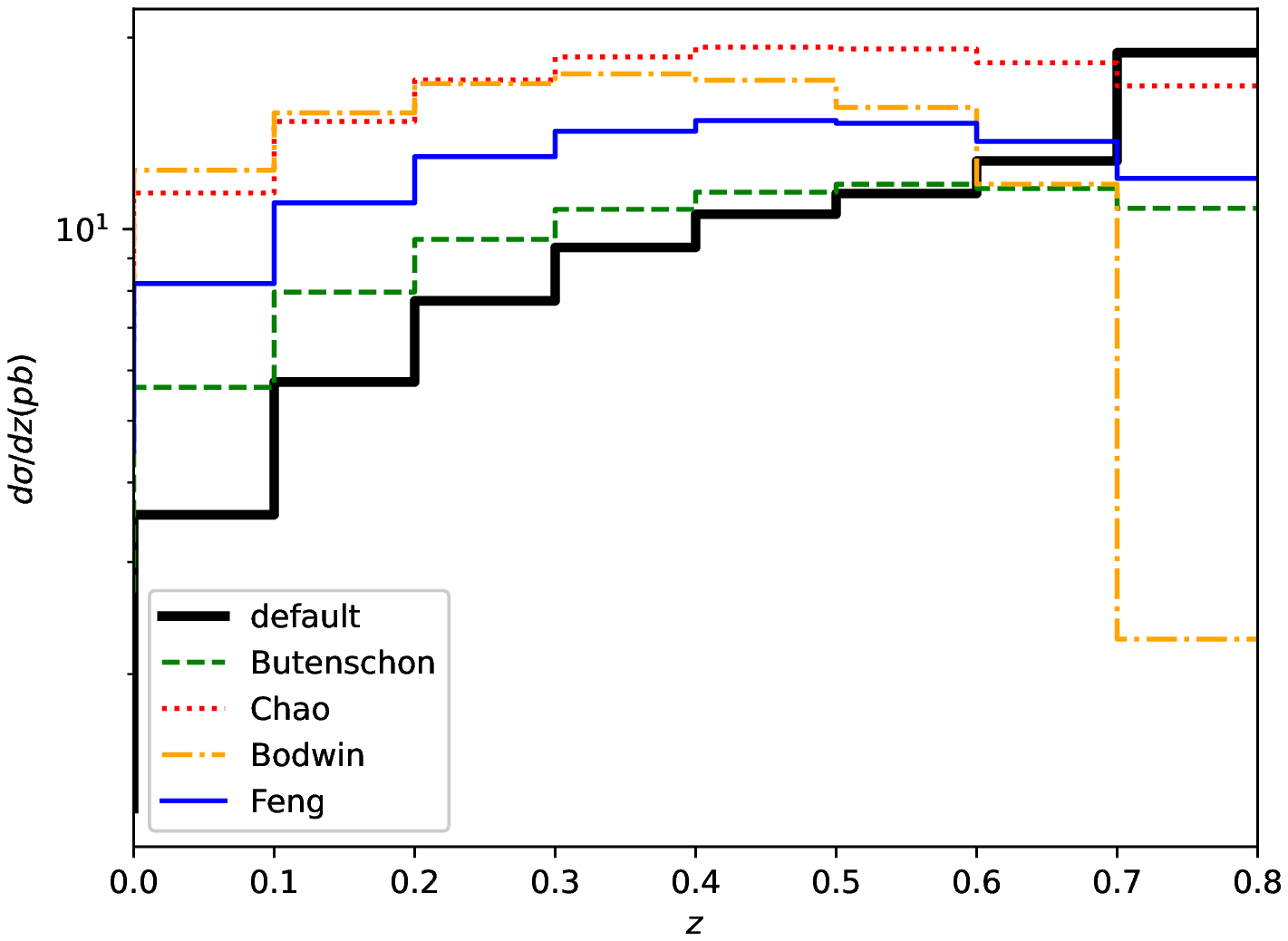}
\caption{\label{fig:ldme_HERA}
The differential cross sections of the $\eta_c$ leptoproduction with respect to $p_t^2$,
$p_t^{\star2}$, $Q^2$, $W$, and $z$ in HERA experimental condition.
}
\end{figure}

\begin{figure}
\includegraphics[scale=0.5]{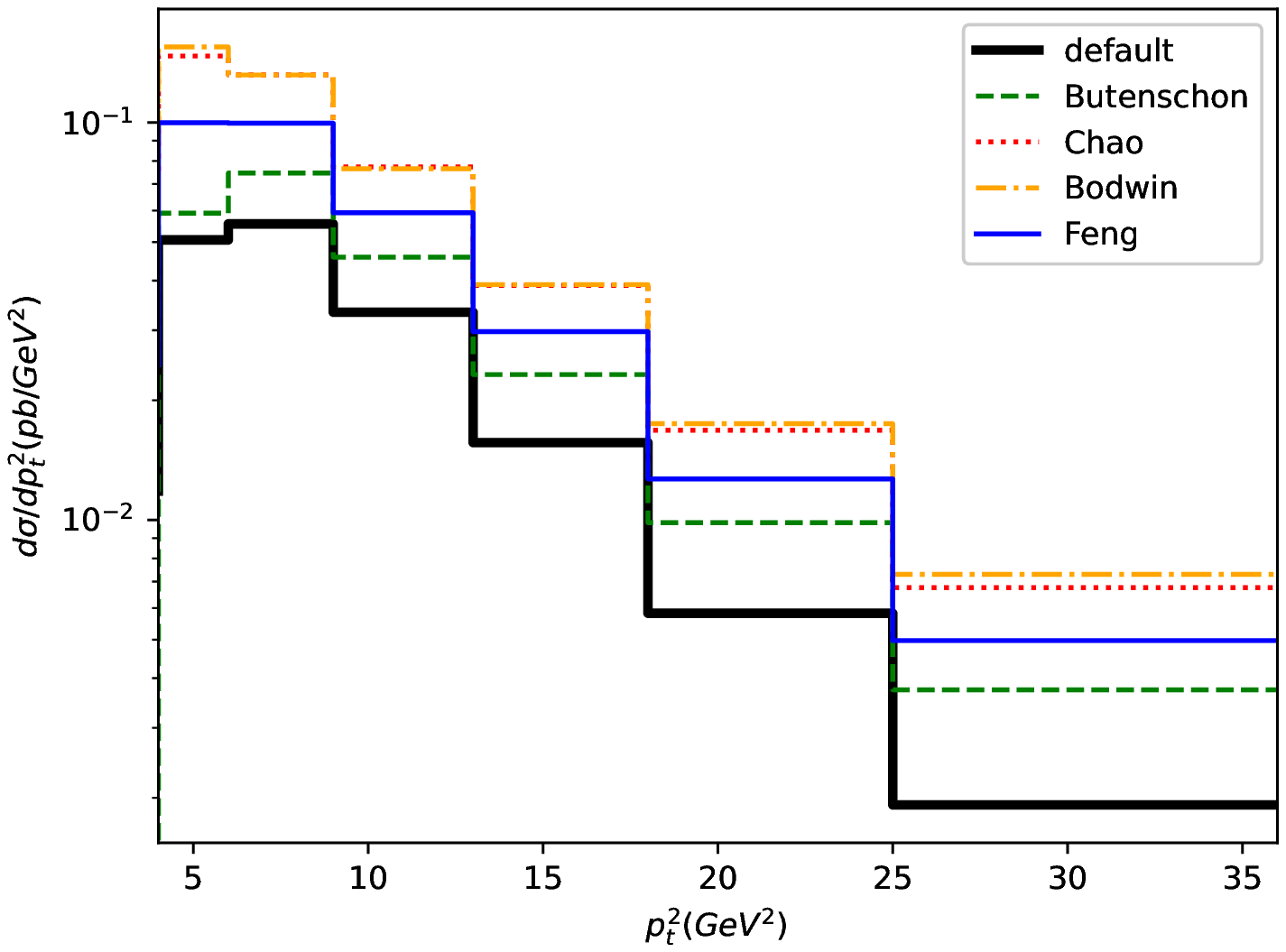}
\includegraphics[scale=0.5]{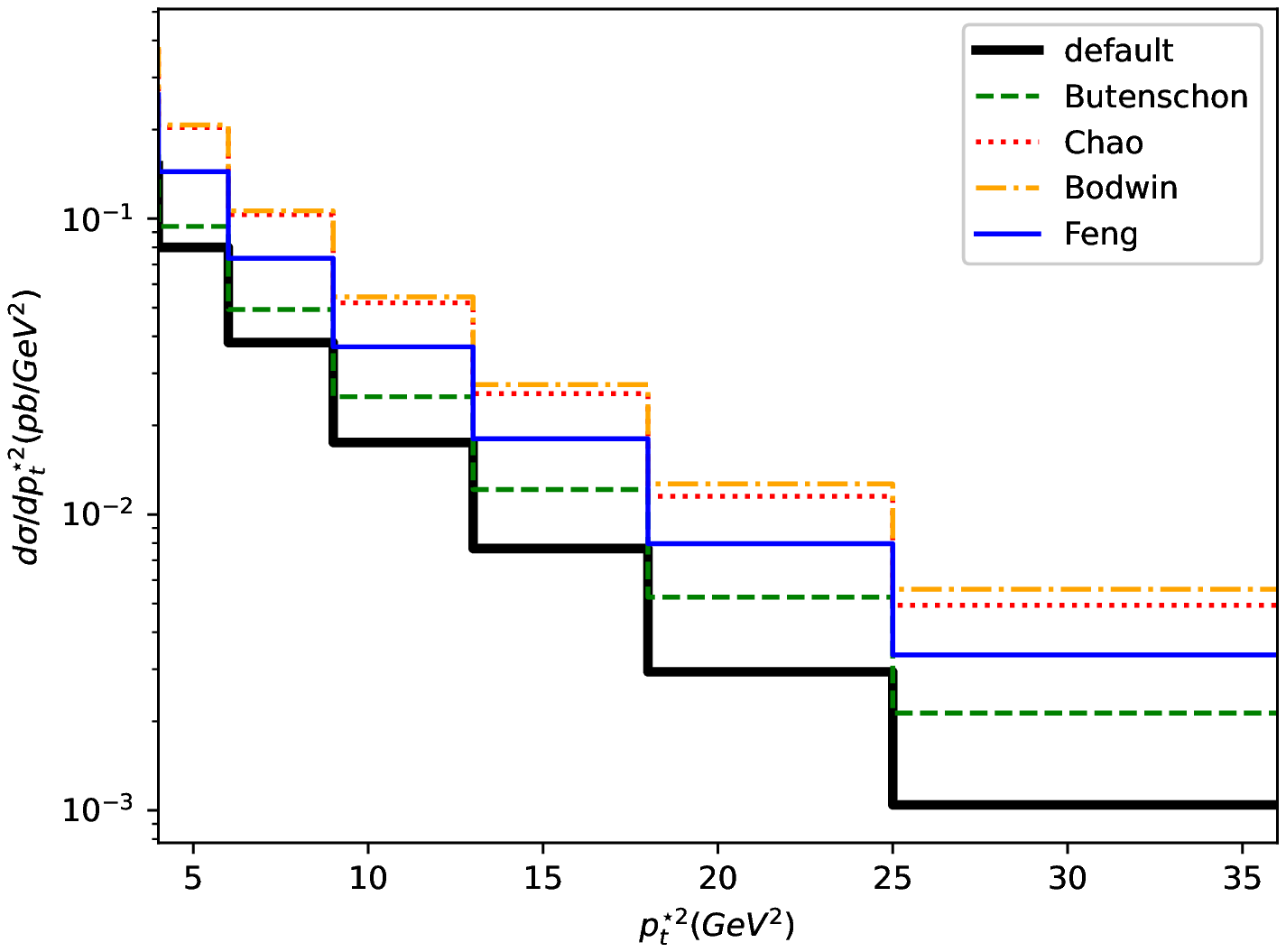}\\
\includegraphics[scale=0.5]{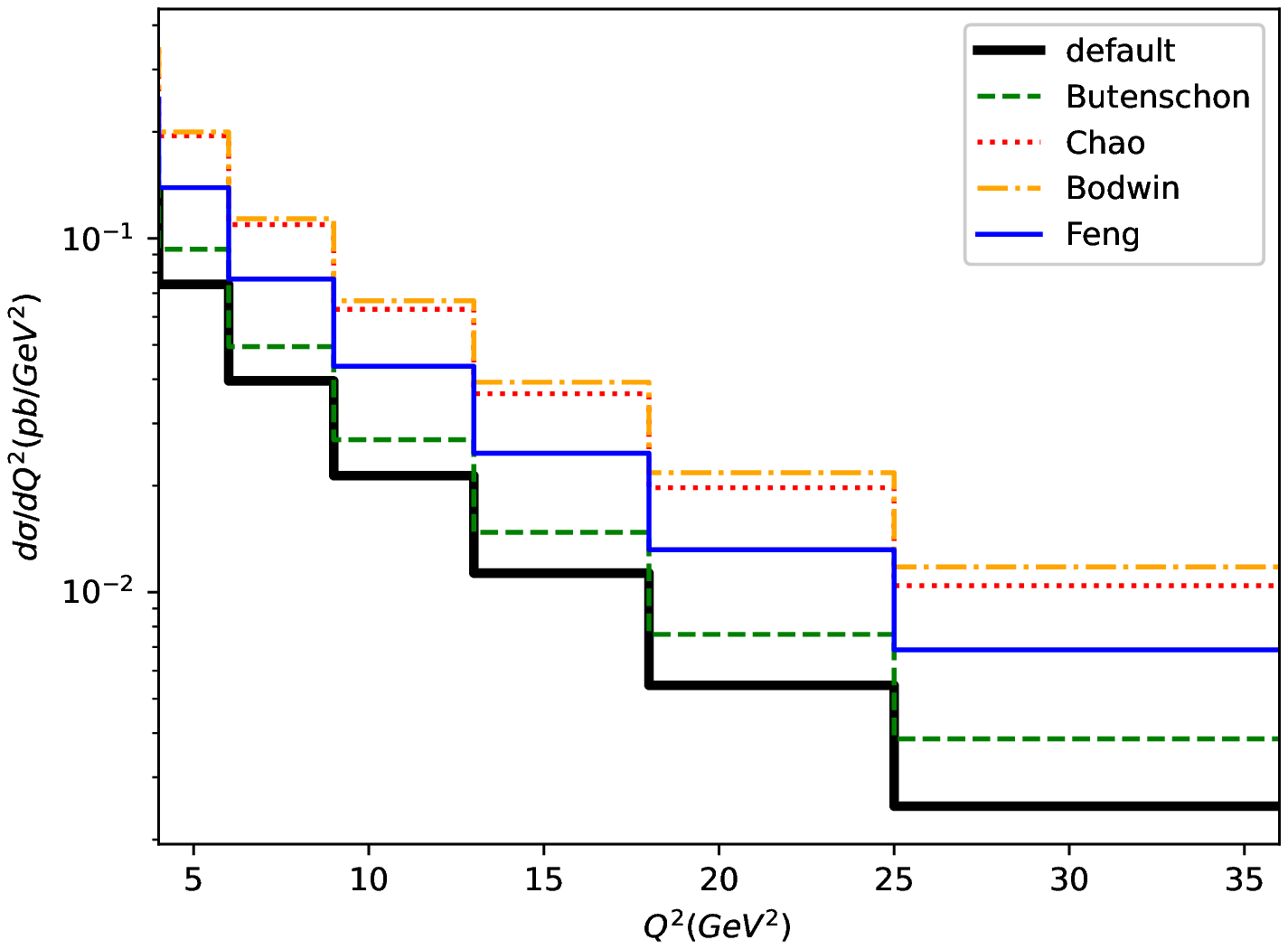}
\includegraphics[scale=0.5]{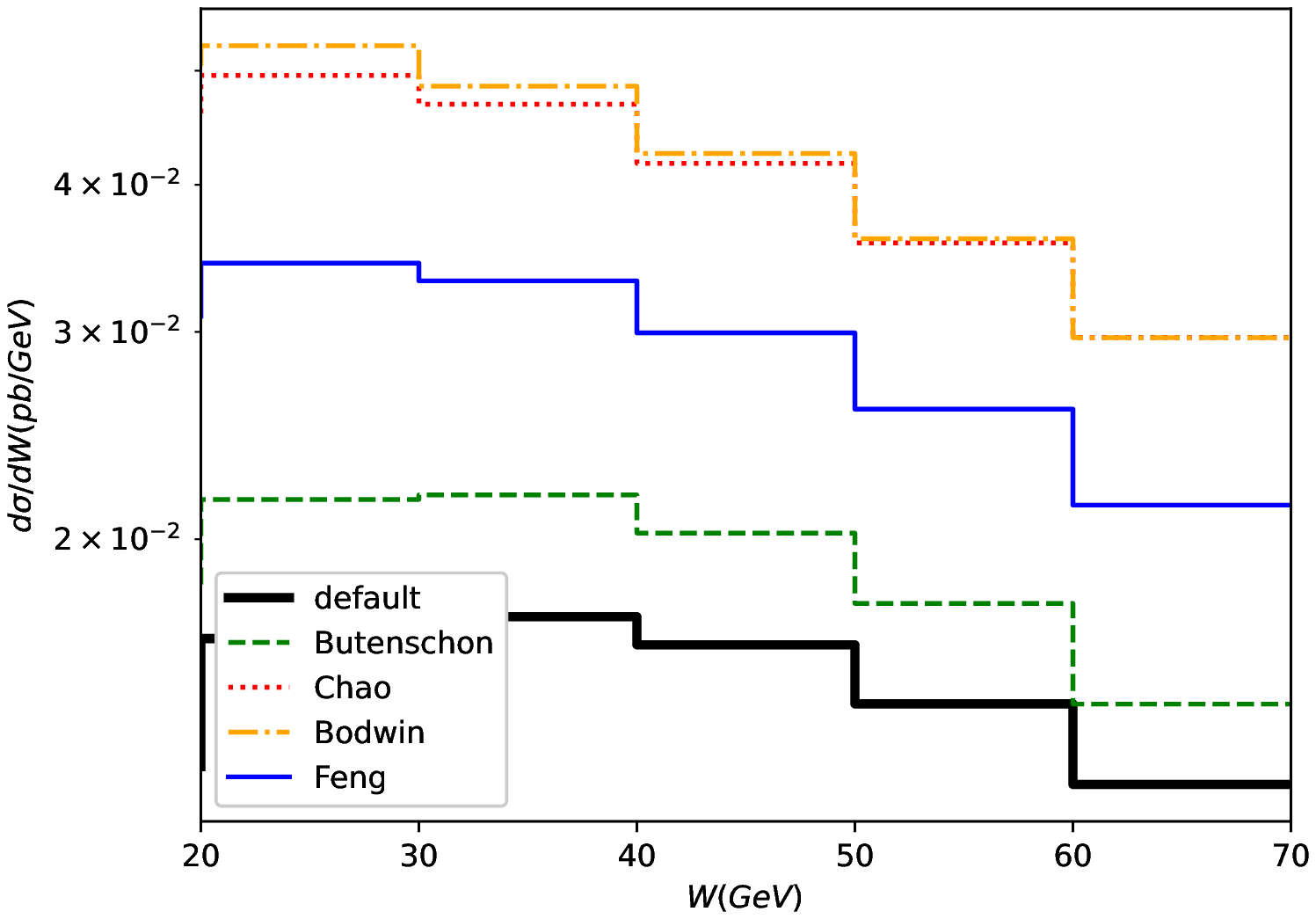}\\
\includegraphics[scale=0.5]{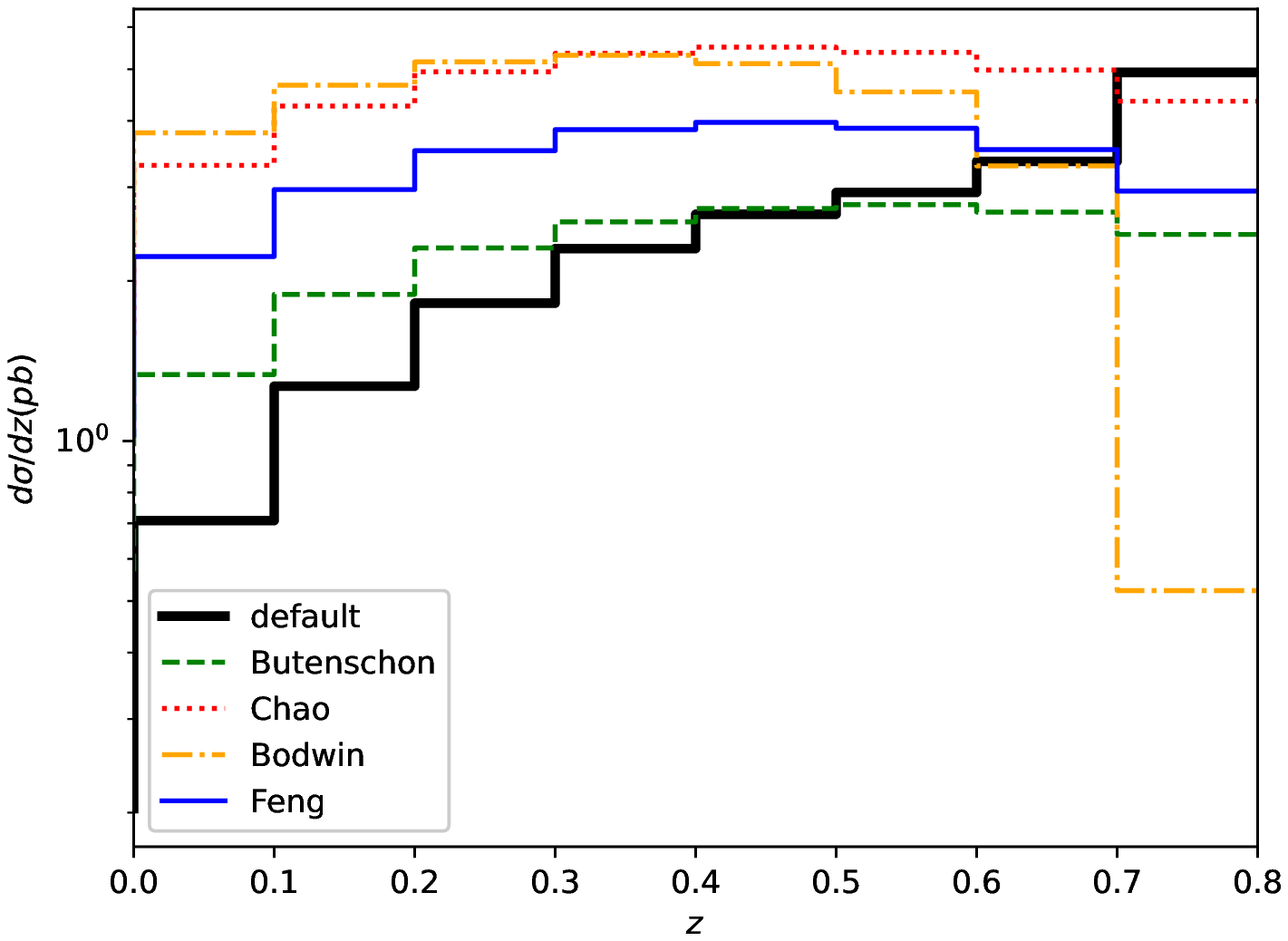}
\caption{\label{fig:ldme_EIC}
The differential cross sections of the $\eta_c$ leptoproduction with respect to $p_t^2$,
$p_t^{\star2}$, $Q^2$, $W$, and $z$ in HERA experimental condition.
}
\end{figure}

In Figure~\ref{fig:ldme_HERA} and Figure~\ref{fig:ldme_EIC},
the differential cross sections with respect to the five different kinematic variables are presented for HERA and EIC experimental conditions, respectively.
Unlike the $J/\psi$ hadroproduction case, the five sets of long-distance matrix elements lead to different results.
This feature has also been observed in the $\eta_c$ photoproduction~\cite{Zhang:2019wxo}.
For all the distributions, the long-distance matrix elements taken from References~\cite{Chao:2012iv, Bodwin:2015iua} give almost the same results,
since both of them are dominated by the $^3S_1$ colour-octet channel.
The other three independent results can be easily distinguished in any of the plots.

\section{Summary\label{sec:summary}}

In this paper, we study the $\eta_c$ leptoproduction at the HERA and the future EIC.
The nonrelativistc-QCD results for this process are obtained for the first time.
Owing to the off-shell initial photon, the calculation of this process is actually very complicated.
Having got the short-distance coefficients for the $c\bar{c}^1S_0^{[1]}$,
$c\bar{c}^1S_0^{[8]}$, $c\bar{c}^3S_1^{[8]}$, and $c\bar{c}^1P_1^{[8]}$ production,
we presented the differential cross sections for the $\eta_c$ leptoproduction with respect to $p_t^2$,
$p_t^{\star2}$, $Q^2$, $W$, and $z$.
Unlike the $\eta_c$ photoproduction, in this process, the contributions from the colour-singlet channel is important,
which is mainly due to a next-to-leading-power term rising from the off-shell initial photon.
The four channels can be well distinguished by studying different kinematic distributions.
By varying the scales ($m_c$, $\mu_r$, and $\mu_f$), we find the cross section is just slight dependent on them.
At the end of this paper, we presented the results for different sets of long-distance matrix elements,
and find that this process can provide a good opportunity to distinguish them.

\acknowledgments

This work is supported by the National Natural Science Foundation of China (Grant No. 11965006).

%\bibliography{paper}% Produces the bibliography via BibTeX.
\providecommand{\href}[2]{#2}\begingroup\raggedright\endgroup

\end{document}